\documentclass[showpacs,
 amsmath,amssymb,nofootinbib,
 reprint,
]{revtex4-1}

\usepackage{graphicx}
\usepackage{dcolumn}
\usepackage{bm}
\usepackage[applemac]{inputenc}
\usepackage[T1]{fontenc}
\usepackage{dsfont}
\usepackage{amsmath}
\usepackage{amssymb}
\usepackage[english]{babel}
\usepackage[applemac]{inputenc}
\usepackage{graphics}
\usepackage{epsfig}
\usepackage{ulem}

\newcommand{\dint}{\text{d}}
\newcommand{\icmpl}{\text{i}}

\begin{document}
\title{Shear Viscosity of a Hot Pion Gas}
\author{Robert Lang}
  \email{robert.lang@ph.tum.de}
  \affiliation{Physik Department, Technische Universit\"{a}t M\"{u}nchen, D-85747 Garching, Germany}
\author{Norbert Kaiser}
  \email{norbert.kaiser@ph.tum.de}
  \affiliation{Physik Department, Technische Universit\"{a}t M\"{u}nchen, D-85747 Garching, Germany}
\author{Wolfram Weise}
  \email{wolfram.weise@ph.tum.de}
  \affiliation{Physik Department, Technische Universit\"{a}t M\"{u}nchen, D-85747 Garching, Germany}
\date{July 16, 2012}

\begin{abstract}
The shear viscosity of an interacting pion gas is studied using the Kubo formalism as a microscopic description of thermal systems close to global equilibrium. We implement the skeleton expansion in order to approximate the retarded correlator of the viscous part of the energy-momentum tensor. After exploring this in $g\phi^4$ theory we show how the skeleton expansion can be consistently applied to pions in chiral perturbation theory. The shear viscosity $\eta$ is determined by the spectral width, or equivalently, the mean free path of pions in the heat bath. We derive a new analytical result for the mean free path which is well-conditioned for numerical evaluation and discuss the temperature and pion-mass dependence of the mean free path and the shear viscosity. The ratio $\eta/s$ of the interacting pion gas exceeds the lower bound $1/4\pi$ from AdS/CFT correspondence. \\
\end{abstract}


\maketitle

\newpage

\section{Introduction}
One of the lessons learned from experiments at the Relativistic Heavy Ion Collider (RHIC) in search for the quark-gluon plasma is the fact that the deconfined quark-gluon matter behaves as an almost-perfect fluid \cite{Hei05, BRAHMSatRHIC05, PHENIXatRHIC05, PHOBOSatRHIC05, STARatRHIC05, Romatschke07, Luzum08}. Nowadays we can look forward to data from the Large Hadron Collider (LHC) at even higher energy densities and temperatures \cite{Armest08, Hei08, Armest08, Aamodt10, Aamodt11, Aamodt12}.

In the literature various perturbative approaches are used for investigating (transport) properties of matter under extreme conditions \cite{NakanoChen07, WangHeinz96, Larionov07, LiuHouLi06, Masaharu08, HaglinPratt94}. In this work we consider an interacting (isospin symmetric) pion gas and focus on its shear viscosity $\eta$. Therewith we approach the properties of hot QCD matter from temperatures below the crossover region from hadronic to partonic matter: $T\lesssim 140\,\text{MeV}$. A standard method to deal with such problems was developed by Kubo relating dissipative quantities to retarded correlators \cite{Kubo57}. We consider a hot pion gas close to thermal equilibrium, so we first perform an expansion in the dissipative forces $\partial^\mu F^\nu$. Then we implement a skeleton expansion of the four-point function entering the thermal viscous correlator $\Pi_\beta(\omega_n)$. The correlations between full propagators are neglected, which results in a quasi-particle approximation. Finally, the chiral 
expansion is used for calculating the pion self energy at finite temperature. We find that the ratio $\eta/s$ (with $s$ the entropy density) decreases as function of temperature but always exceeds the AdS/CFT bound $1/4\pi$.

\section{Relativistic Hydrodynamics}
We start with a macroscopic view on dissipative fluids. By definition we are dealing with a fluid that does not fulfill Pascal's law, i.e. its pressure exerts transverse surface forces. Classically the dynamics of such a fluid is described by the Navier-Stokes equation. In relativistic hydrodynamics one extends the energy-momentum tensor of a perfect fluid by the dissipative tensor $\tau^{\mu\nu}$. With the Lorentz factor $1/\gamma(x)=\sqrt{1-\vec{v}\,^2(x)}$ let $u^\mu(x)=\gamma(x)(1,\vec{v}(x))$ denote the four-velocity, $\epsilon(x)$ the energy density and $P(x)$ the pressure. Then the energy-momentum tensor reads
\begin{equation}
 \label{EnergyMomentumTensorDiss}
  T^{\mu\nu}=u^\mu u^\nu(\epsilon+P)-Pg^{\mu\nu}+\tau^{\mu\nu}\,.
\end{equation}
The form of the dissipative tensor $\tau^{\mu\nu}$ can be fixed by the following three conditions \cite{Yagi08,Wei72}: $u_\mu \tau^{\mu\nu}=0$ (a direct consequence of \eqref{EnergyMomentumTensorDiss} when evaluated in the rest frame), $\partial_\mu s^\mu\geq 0$ (second law of thermodynamics with the entropy-density current $s^\mu=su^\mu$), and the assumption that only first-order derivatives $\partial^\mu u^\nu$ contribute to $\tau^{\mu\nu}$. In this case the dissipative tensor is parametrized by the shear and bulk viscosities, $\eta$ and $\zeta$, respectively:
\begin{equation}
 \label{DissTensorGeneral}
  \tau^{\mu\nu}=\eta\big[\partial_\bot^\mu u^\nu+\partial_\bot^\nu u^\mu -\frac 23\Delta^{\mu\nu}(\partial_\bot\cdot\, u)\big]+\zeta\Delta^{\mu\nu}(\partial_\bot\cdot\, u)\, ,
\end{equation}
where we have introduced the following notations
\begin{equation}
 \label{DissTensorAbbrev}
 \Delta^{\mu\nu}= g^{\mu\nu}-u^\mu u^\nu, \;\;\; \partial_\bot^\mu= \partial^\mu-u^\mu(u\cdot\partial)=\Delta^{\mu\nu}\partial_\nu\,.
\end{equation}
The second law of thermodynamics ensures that both transport coefficients are non-negative: $\eta,\zeta\geq 0$. We concentrate on the shear viscosity which describes the traceless part of $\tau^{\mu\nu}$. A more detailed discussion of the hydrodynamical properties with inclusion of shear and bulk viscosity can be found in Refs. \cite{Muronga02,Yagi08}.

Relativistic hydrodynamics is applicable to dissipative systems close to global thermal equilibrium. Experiments at RHIC suggest that the hot matter produced in heavy-ion collisions reaches local equilibrium after a short thermalization time, $\tau_0\lesssim 1\,\text{fm}$ \cite{Hei05, BRAHMSatRHIC05, PHENIXatRHIC05, PHOBOSatRHIC05, STARatRHIC05, Romatschke07, Luzum08, Bjorken83}. A system in local equilibrium is characterized by the fact that one can divide the macroscopic system into mesoscopic zones in such a way that each zone by itself is in thermal equilibrium.

The more dissipative the system is, the more it deviates from global equilibrium: four-velocity, temperature, energy density, pressure, etc. display stronger space-time variations. The concept of local equilibrium therefore translates into small values of gradients, e.g. $\partial^\mu u^\nu$. Therefore second-order terms can be neglected as already done in the parametrization of $\tau^{\mu\nu}$ in Eq.~\eqref{DissTensorGeneral}. This approximation is justified when the system is still dissipative but close to global equilibrium. The matter produced at RHIC meets this condition, after thermalization.

\section{Non-Equilibrium Thermodynamics}
The shear and bulk viscosities, $\eta$ and $\zeta$, are macroscopic parameters for the dissipative part, Eq.~\eqref{DissTensorGeneral}, of the energy-momentum tensor. Now we recall the Kubo-type formula for the shear viscosity following Zubarev's approach \cite{Hosoya84,Zubarev74}. First we introduce a statistical operator in the Schr\"{o}dinger picture:
\begin{equation}
 \label{DefStatOpNonEq}
  \rho=\frac 1Q\exp{\left[-\int\dint^3x\;\mathcal{B}(\vec{x})\right]}\, ,
\end{equation}
where $Q=\text{Tr}\,\exp{\left[-\int\dint^3x \;\mathcal{B}(\vec{x})\right]}$ ensures $\text{Tr}\,\rho=1$. We have defined the \textit{time-independent} operator,
\begin{equation}
\label{DefB}
\mathcal{B}(\vec{x})=F^\mu(\vec{x},t)T_{0\mu}(\vec{x},t)-\int_{-\infty}^t\dint t'\;T_{\mu\nu}(\vec{x},t')\partial^\mu F^\nu(\vec{x},t')\,,
\end{equation}
with $\partial_t\mathcal{B}(\vec{x})=0$, and the four-vector
\begin{equation}
 \label{DefFourVectorF}
  F^\mu(\vec{x},t)=\beta_\text{s}(\vec{x},t)u^\mu(\vec{x},t)\,.
\end{equation}
Here we have introduced the inverse \textit{proper temperature},
\begin{equation}
  \label{DefBetas}
  \begin{aligned}
   \beta_\text{s}=\frac{1}{T_\text{s}}=\frac{\gamma}{T}\,,
  \end{aligned}
\end{equation}
again with the Lorentz factor $\gamma$. Evaluating $\beta_\text{s}^{-1}$ in the local rest frame (the distinguished frame of the heat bath), one recovers the common temperature $T$. Note that with $\beta_\text{s}$ a Lorentz scalar, $F^\mu$ is indeed a four-vector. The operator $\mathcal{B}$ is Lorentz \textit{variant}, but with the additional spatial integration, the statistical operator $\rho$ is indeed a Lorentz scalar. With $\beta=1/T$ being the inverse temperature, it can be written as
\begin{equation}
 \label{StatOpNonEqShaped}
  \rho = \frac 1Q\,\exp[-\beta H+\mathcal{A}]\,,
\end{equation}
where $\beta H=\int\dint^3x\;F^\mu(\vec{x},t)T_{0\mu}(\vec{x},t)$ and $\mathcal{A}$ denotes the spatial integral over the second part of $\mathcal{B}(\vec{x})$ in Eq.~\eqref{DefB}.
                                                                                                                                                                                                                                                                                                                                                                                                                                                                                                                                                                         
The statistical operator $\rho$ in Eq.~\eqref{StatOpNonEqShaped} provides one possible approach to non-equilibrium thermodynamics. It is not mandatory to introduce an operator $\mathcal{B}$ as done in \eqref{DefB}, but the resulting form for $\rho$ is physically meaningful: first because it depends on the \textit{dissipative forces} $\partial^\mu F^\nu$, a measure for deviations from global equilibrium. Secondly, for vanishing $\mathcal{A}$, it reproduces the statistical operator for equilibrium systems, $\rho_0=\left.\rho\right|_{\mathcal{A}=0}$\,. Note that $H$ is not a free Hamiltonian but represents a fully interacting theory. We mention other methods, e.g. the Boltzmann equation, for investigating non-equilibrium systems \cite{NakanoChen07,Larionov07}.

The following quantities are introduced in order to decompose a given energy-momentum tensor $T^{\mu\nu}$:
\begin{equation}
\label{Def4Coefficients}
\begin{aligned}
\epsilon &= u_\mu u_\nu T^{\mu\nu}\,, \;\;\;\;\;\;\;\;\; P=-\frac 13\Delta_{\mu\nu}T^{\mu\nu}\, , \\
K_\mu &= \Delta_{\mu\rho}u_\sigma T^{\rho\sigma}\,, \;\;\; \pi_{\mu\nu}= \big(\Delta_{\mu\nu}\Delta_{\rho\sigma}-\frac 13\Delta_{\mu\rho}\Delta_{\nu\sigma}\big)T^{\rho\sigma}\,,
\end{aligned}
\end{equation}
with $\Delta^{\mu\nu}$ given in Eq.~\eqref{DissTensorAbbrev}, so that:
\begin{equation}
  \label{TmunuDecomp4}
  T^{\mu\nu}=\epsilon\, u^\mu u^\nu -P\Delta^{\mu\nu}+K^\mu u^\nu+K^\nu u^\mu+\pi^{\mu\nu}\, .
\end{equation}
Within linear response theory one can derive the impact of the dissipative forces on the energy-momentum tensor. We assume these forces to be small compared to typical energies of the system: $\langle\beta H\rangle_0\gg\langle\mathcal{A}\rangle_0$, where $\langle\cdot\rangle_0$ denotes the thermal expectation value with respect to the equilibrium statistical operator $\rho_0$, i.e. $\langle\cdot\rangle_0=\text{Tr}\left(\cdot\,\rho_0\right)$. Up to linear order in the dissipative forces the statistical operator in Eq.~\eqref{StatOpNonEqShaped} reads
\begin{equation}
  \label{StatOperLinearResp}
  \rho\approx\rho_0\left(1+\int_0^1\dint\xi\;\text{e}^{\beta H\xi}\mathcal{A}\,\text{e}^{-\beta H\xi}-\langle\mathcal{A}\rangle_0\right).
\end{equation}
In particular, the linear response of the microscopic \textit{viscous-stress tensor} $\pi^{\mu\nu}$ to dissipative forces leads to a connection between its correlation function and the macroscopic shear-viscosity parameter:
\begin{equation}
  \label{EtaPiCorrNonRet}
  \eta(\vec{x},t)=\frac{\beta_\text{s}}{10}\int\dint^3x'\,\int_{-\infty}^t\dint t'\;\big(\pi^{\mu\nu}(\vec{x},t),\pi_{\mu\nu}(\vec{x}\,',t')\big),
\end{equation}
where the structure of the correlator $(X,Y)$ follows from the statistical operator \eqref{StatOperLinearResp}:
\begin{equation}
  \label{DefCorrNonRet}
\big(X,Y\big)=\frac{1}{\beta}\int_0^\beta\dint\tau\;\big\langle X\left[\text{e}^{H\tau}Y\,\text{e}^{-H\tau}-\langle Y\rangle_0\right]\big\rangle_0\;.
\end{equation}
One can express this correlator as a real-time integral over a retarded correlator:
\begin{equation}
  \label{TransCorrRetNonRet}
  \big(\pi^{\mu\nu}(x),\pi_{\mu\nu}(\vec{x}\,',t')\big) \approx -\frac{1}{\beta}\int_{-\infty}^{t'}\dint\tilde{t}\,\langle\pi^{\mu\nu}(x),\pi_{\mu\nu}(\vec{x}\,',\tilde{t}\,)\rangle_\text{R}\,,
\end{equation}
where $x=(\vec{x},t)$ and the retarded correlator is defined as
\begin{equation}
  \label{DefRetCorrelator}
  \langle X(t),Y(t')\rangle_{\text{R}}=-\icmpl\,\theta(t-t')\langle\left[X(t),Y(t')\right]\rangle_0\;.
\end{equation}
The approximation \eqref{TransCorrRetNonRet} becomes exact in the large-time limit $t\to\infty$, when the system reaches global equilibrium. Finally, when combining Eqs.~\eqref{EtaPiCorrNonRet} and \eqref{TransCorrRetNonRet}, and evaluating at the local rest frame ($\beta_\text{s}/\beta\to 1$), one finds the Kubo-type formula for the shear-viscosity field:
\begin{equation}
  \label{ShearViscosity}
  \eta(\vec{x},t)=-\frac{1}{10} \int\dint^3x '\int_{-\infty}^t \!\!\!\!\!\dint t'\int_{-\infty}^{t'}\!\!\!\!\!\dint\tilde{t}\;\langle\pi^{\mu\nu}(\vec{x},t),\pi_{\mu\nu}(\vec{x}\,',\tilde{t})\rangle_{\text{R}}\,.
\end{equation}
Details concerning this derivation of $\eta$ are worked out in Ref. \cite{Hosoya84}. For completeness we also give the expressions for the bulk viscosity $\zeta$ and the heat conductivity $\kappa$:
\begin{equation}
  \label{BulkViscosity}
    \zeta(\vec{x},t) =-\int\dint^3x '\int_{-\infty}^t \!\!\!\!\!\dint t'\int_{-\infty}^{t'} \!\!\!\!\!\dint\tilde{t}\;\langle P'(\vec{x},t),P'(\vec{x}\,',\tilde{t}\;)\rangle_{\text{R}}\, ,
\end{equation}
\begin{equation}
  \label{HeatKappa}
  \kappa(\vec{x},t) = \frac 13 \int\dint^3x ' \int_{-\infty}^t \!\!\!\!\!\dint t' \int_{-\infty}^{t'} \!\!\!\!\!\dint\tilde{t}\;\langle K_\mu(\vec{x},t),K^\mu(\vec{x}\,',\tilde{t}\;)\rangle_{\text{R}}\, ,
\end{equation}
where $P'=\epsilon\,\frac{\partial\langle P\rangle_0}{\partial\langle\epsilon\rangle_0}-P$. The heat conductivity can be neglected \cite{Danielewicz85} if the chemical potentials are small compared to the temperature, i.e. if $\mu_j/T\ll 1$.

\section{General Perturbative Evaluation of the Shear Viscosity}
\subsection{Skeleton Expansion}
In this section we show how an expansion of the four-point correlator in \textit{full} Matsubara propagators offers a suitable method for calculating the shear viscosity. The conceptional framework for this expansion is first worked out in $g\phi^4$ theory:
\begin{equation}
 \label{LagrangePhi4Bare}
  \begin{aligned}
  \mathcal L=\frac 12(\partial_\mu\phi)(\partial^\mu\phi)-\frac 12 m^2\phi^2-g\phi^4\, ,
  \end{aligned}
\end{equation}
where $\phi$ is a real scalar field, $m$ denotes the bare particle mass and $g>0$ is a small coupling constant. The viscous-stress tensor is determined entirely by the momentum-dependent parts of the Lagrangian:
\begin{equation}
\label{piLagrangePhi4Bare}
\pi_{\mu\nu} =(\Delta_\mu^\rho\Delta_\nu^\sigma-\frac 13\Delta_{\mu\nu}\Delta^{\rho\sigma})(\partial_\rho\phi)(\partial_\sigma\phi)\, .
\end{equation}
In general, the shear viscosity and other quantities such as pressure, temperature, four-velocity, etc.~are fields, but on large time scales, when local equilibrium approaches global equilibrium, they reduce to temperature-dependent numbers. We evaluate the shear viscosity \eqref{ShearViscosity} at the origin of Minkowski space:
\begin{equation}
\label{EtaZero}
  \eta(0)=-\frac{1}{10}\int_{-\infty}^{0}\dint t'\int_{-\infty}^{t'}\dint\tilde{t}\; \Pi_\text{R}(\tilde{t}\,)\, ,
\end{equation}
with the spatially-integrated retarded Green's function for the viscous-stress tensor
\begin{equation}
  \Pi_\text{R}(\tilde{t}\,)=-\icmpl\int\dint^3x'\;\left\langle\left[ \pi^{\mu\nu}(0),\pi_{\mu\nu}(\vec{x}\,',\tilde{t}\,)\right]\right\rangle_0.
\end{equation}
Using the analytical continuation to Minkowski space via ${\text i}\omega_n\mapsto p_0+\icmpl\epsilon$, we need to calculate the spatially-integrated \textit{thermal} Green's function, labeled by $\beta$:
\begin{equation}
\Pi_\beta(\tau)=\int\dint^3x'\; \left\langle \mathcal{T}_\tau\left[\pi^{\mu\nu}(0)\pi_{\mu\nu}(\vec{x}\,',\tau)\right]\right\rangle_0\,,
\end{equation}
where $\mathcal{T}_\tau$ stands for the time-ordering prescription. Consider next the Fourier transform of $\Pi_\beta(\tau)$:
\begin{widetext}
\begin{equation}
 \label{spatIntThermGreen}
  \begin{aligned}
  \Pi_\beta(\omega_n)=\frac{1}{V^2}\int_{0}^\beta\dint\tau\; \text{e}^{\icmpl\omega_n\tau}\int\frac{\dint^3p}{(2\pi)^3}\; \big(\Delta^{\mu\rho}\Delta^{\nu\sigma}-\frac 13\Delta^{\mu\nu}\Delta^{\rho\sigma}\big)p_\mu p_\nu p_\rho p_\sigma \left\langle \mathcal{T}_\tau\left[\phi(0)\phi(0)\phi(\vec{p},\tau)\phi(-\vec{p},\tau)\right]\right\rangle_0\;,
  \end{aligned}
\end{equation}
\end{widetext}
where $\phi(\vec{p},\tau)=\int\dint^3x\; \text{e}^{\icmpl\vec{p}\cdot\vec{x}} \phi(\vec{x},\tau)$. The volume factors come from residual Fourier integrals originating from the normalization $V\int\dint^3p=(2\pi)^3$. They ensure the correct mass dimension of the momentum-integrated thermal Green's function: $\text{dim}\,\Pi_\beta(\omega_n)=4$, since $\text{dim}\,\phi(\vec{p},\tau)=-2$. The periodic boundary conditions of the bosonic fields in imaginary time are realized by the compact interval $\tau\in[0,\beta]$ in Eq.~\eqref{spatIntThermGreen}.

Consider now the four-point correlation function $\left\langle T_\tau\left[\phi(0)\phi(0)\phi(\vec{p},\tau)\phi(-\vec{p},\tau)\right]\right\rangle_0$ in detail. Let us recall the covariance of two random variables $X$ and $Y$:
\begin{equation}
 \label{correlationDefXY}
  \begin{aligned}
  \text{Cov}(X,Y)=\big\langle\left(X\!-\!\langle X\rangle\right)\left(Y\!-\!\langle Y\rangle\right)\big\rangle=\langle XY\rangle\!-\!\langle X\rangle\langle Y\rangle\,.
  \end{aligned}
\end{equation}
It demonstrates that the expectation value of a product factorizes if and only if the random variables are uncorrelated. In this sense we introduce the \textit{skeleton expansion} as an expansion in \textit{full propagators} subject to correlations \textit{between} them treated perturbatively \cite{Hung92,Hosoya84}. At leading order, when neglecting these correlations, one obtains
\begin{equation}
\label{correlation4point}
\begin{aligned}
   & \frac{1}{V^2}\left\langle \mathcal{T}_\tau\left[\phi(0)\phi(0)\phi(\vec{p},\tau)\phi(-\vec{p},\tau)\right]\right\rangle_0 \\
& \approx \frac{2}{V^2}\left\langle \mathcal{T}_\tau\left[\phi(0)\phi(\vec{p},\tau)\right]\right\rangle_0 \left\langle \mathcal{T}_\tau\left[\phi(0)\phi(-\vec{p},\tau)\right]\right\rangle_0\\
  &\hspace{0.5cm} +\frac{1}{V^2}\left\langle \mathcal{T}_\tau\left[\phi(0)\phi(0)\right]\right\rangle_0 \left\langle \mathcal{T}_\tau\left[\phi(\vec{p},\tau)\phi(-\vec{p},\tau)\right]\right\rangle_0 \\
  &\approx\frac{2}{V^2}\left(\left\langle \mathcal{T}_\tau\left[\phi(0)\phi(\vec{p},\tau)\right]\right\rangle_0 \right)^2 = 2\, G^2_\beta(\vec{p},\tau)\, .
  \end{aligned}
\end{equation}
The first approximation comes from forming all possible contractions that lead to full propagators (two-point functions) and factorizing them. The second one comes from neglecting the vacuum loops. Moreover the bosonic Matsubara propagator does not dependent on the direction of the momentum flow: $G_\beta(\vec{p},
\tau)=G_\beta(-\vec{p},\tau)$. The skeleton expansion in $g\phi^4$ theory can be represented up to next-to-leading order by the following diagrams:
\begin{equation}
\label{feySkeletonLONLO}
\frac{1}{V^2}\left\langle \mathcal{T}_\tau\left[\phi^2(0)\;\phi^2(\vec{p},\tau)\right]\right\rangle_0 \approx
\;\parbox{0.2\linewidth}{\includegraphics[width=\linewidth]{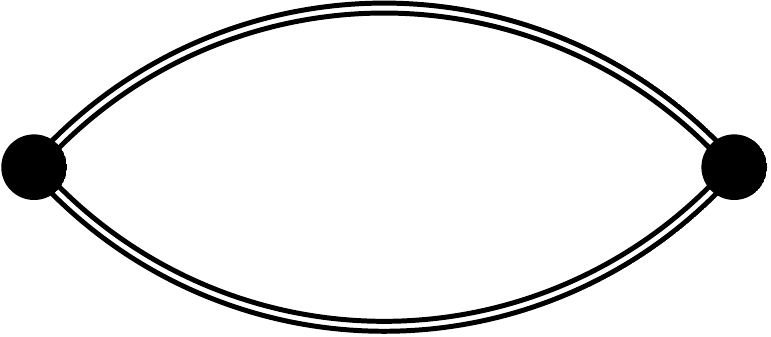}}\; + \; \parbox{0.2\linewidth}{\includegraphics[width=\linewidth]{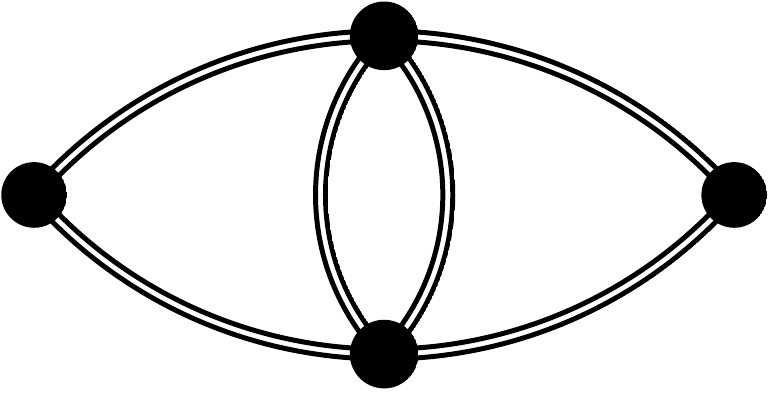}}\;.
\end{equation}
Here, the lines denote \textit{full} bosonic Matsubara propagators with self-energy insertions included. In the following calculation we restrict ourselves to the  one-loop diagram since the primary application will be focused on pions within chiral perturbation theory. However, in scalar $g\phi^4$ theory a resummation of ladder diagrams \cite{JeonSkeleton1995, JeonYaffeSkeleton1996} is necessary to obtain the correct leading-order result for small coupling constants $g\ll 1$. We briefly discuss this resummation and its impact on chiral perturbation theory in Appendix \ref{AppResum}.

Evaluating Eq.~\eqref{feySkeletonLONLO} at one-loop level leads to
\begin{equation}
\label{LOskeletonTemporal}
\Pi_\beta(\omega_n)=2\int_{0}^\beta\dint\tau\; \text{e}^{\icmpl\omega_n\tau}\int \frac{\dint^3p}{(2\pi)^3}\; p^{(\mu}\,p^{\nu)} p_{(\mu}\,p_{\nu)} G^2_\beta(\vec{p},\tau)\,,
\end{equation}
where we have abbreviated the contraction of $p^4$ in Eq.~\eqref{spatIntThermGreen} by $p^{(\mu}\,p^{\nu)} p_{(\mu}\,p_{\nu)}$.

It is convenient to express the Matsubara propagator in a spectral representation,
\begin{equation}
 \label{temporalSpatialThermalGreenTauBoth}
  G_\beta(\vec{p},\tau)=\int_{-\infty}^{\infty}\dint\omega\;\rho(\omega,\vec{p}\,)\, n(\omega)\, \text{e}^{\omega|\tau|}\,,
\end{equation}
with the periodicity $G_\beta(\vec{p},\tau)=G_\beta(\vec{p},\tau-\beta)$. Here $n(\omega)=(\text{e}^{\beta\omega}-1)^{-1}$ denotes the Bose distribution and $\rho(\omega,\vec{p}\,)$ the spectral function.

On the other hand the retarded full propagator including the complex self-energy $\Sigma_\text{R}$ reads:
\begin{equation}
\label{DefFullRetProp}
\begin{aligned}
  G_{\text{R}}^{-1}(p) &=p^2-m_0^2-\text{Re}\,\Sigma_{\text{R}}(p)-\icmpl\,\text{Im}\,\Sigma_{\text{R}}(p) \\
    &\approx (p_0+\icmpl\gamma(p))^2-E_p^2\,,
\end{aligned}
\end{equation}
where we have introduced the renormalized energy $E_p=\sqrt{\vec{p}\,^2+m_0^2+\text{Re}\,\Sigma_{\text{R}}(p)}$ and the spectral (half) width
\begin{equation}
 \label{DefSpectralWidth}
  \gamma(p)=\frac{1}{2}\,\Gamma(p)=-\frac{1}{2p_0}\,\text{Im}\,\Sigma_{\text{R}}(p)\;.
\end{equation}
The spectral function $\rho(p)=\rho(p_0,\vec{p}\,)$ is determined by the imaginary part of the full retarded propagator,
\begin{equation}
\label{SpecImretFullPropagatorAnalyt}
\rho(p)=\frac{1}{2\pi\icmpl}\left[\frac{1}{\left(p_0-\icmpl\gamma(p)\right)^2-E_p^2}-\frac{1}{\left(p_0+\icmpl\gamma(p)\right)^2-E_p^2}\right],
\end{equation}
involving the difference between advanced and retarded propagators. We have used the identity $ -\pi\rho(p)=\text{Im}\, G_{\text{R}}(p)$.

The spectral function \eqref{SpecImretFullPropagatorAnalyt} can be written also in a modified Breit-Wigner form:
\begin{equation}
  \label{SpectralDensityBreitWigner}
  \rho(p)=\frac{1}{\pi}\frac{2p_0\gamma(p)}{(p_0^2-\gamma^2(p)-E_p^2)^2+4p_0^2\gamma^2(p)}\;.
\end{equation}
In this respect our treatment is equivalent to a quasi-particle approximation.

~\\~
\subsection{Analytical Continuation}
Substituting the Matsubara propagator \eqref{temporalSpatialThermalGreenTauBoth} into \eqref{LOskeletonTemporal}, the Green's function of the viscous-stress tensor can be expressed in a spectral representation as well. Using the temporal periodicity of $G_\beta(\vec{p},\tau)$, the $\tau$-integration with a subsequent analytical continuation from discrete Matsubara frequencies, $\omega_n$, to continuous energies, $p_0$, via $\icmpl\omega_n\mapsto p_0+\icmpl\epsilon$ leads to
\begin{widetext}
\begin{equation}
 \label{PiRetInsertSpecFct}
  \begin{aligned}
  \Pi_{\text{R}}(p_0)&=-2\int\frac{\dint^3p}{(2\pi)^3}\; \; p^{(\mu}\,p^{\nu)} p_{(\mu}\,p_{\nu)} \int_{-\infty}^\infty\dint\omega_1 \int_{-\infty}^\infty \dint\omega_2\; \rho(\omega_1,\vec{p}\,)\rho(\omega_2,\vec{p}\,)\;n(\omega_1)n(\omega_2)\;W_\epsilon(\omega_{12},p_0)\, ,
  \end{aligned}
\end{equation}
\end{widetext}
where we have introduced the sum $\omega_{12}=\omega_1+\omega_2$, and the $p_0$-dependent factor $W_\epsilon$:
\begin{equation}
 \label{PiRetDefUVW}
  W_\epsilon(\omega_{12},p_0) =\frac{1}{p_0+\icmpl\epsilon-\omega_{12}} -\frac{1}{p_0+\icmpl\epsilon+\omega_{12}}\;.
\end{equation}
The real part of $W_\epsilon$ is symmetric in $p_0$, 
\begin{equation}
  \text{Re}\,W_\epsilon(-p_0) =\text{Re}\,W_\epsilon(p_0)\;,
\end{equation}
while the imaginary part changes its pole structure under this reflection:
\begin{equation}
  \text{Im}\,W_\epsilon(-p_0) =\text{Im}\,W_{-\epsilon}(p_0)\;.
\end{equation}
This crucial property of $W_\epsilon$ guarantees a non-zero value of the shear-viscosity \eqref{EtaZero}:
\begin{equation}
 \label{EtaZeroRetarded}
  \begin{aligned}
  \eta &=-\frac{1}{10}\int_{-\infty}^0\dint t'\int_{-\infty}^{t'}\dint \tilde{t}\int_{-\infty}^{\infty}\frac{\dint p_0}{2\pi}\;\text{e}^{-\icmpl p_0\tilde{t}}\;\Pi_\text{R}(p_0)\\
  &=-\frac{1}{10}\int_{-\infty}^0\dint\tilde{t}\int_{\tilde{t}}^{0}\dint t'\int_{-\infty}^{\infty}\frac{\dint p_0}{2\pi}\;\text{e}^{-\icmpl p_0\tilde{t}}\;\Pi_\text{R}(p_0)\\
  &=\frac{\icmpl}{10}\left.\frac{\dint}{\dint p_0}\;\Pi_{\text{R}}(p_0)\right|_{p_0=0}\,,
  \end{aligned}
\end{equation}
where we have first interchanged the order of the integration $(\tilde{t},t')$ and then used the functional identity
\begin{equation}
  \int_{-\infty}^0\dint\tilde{t}\int_{\tilde{t}}^0\dint t'\; \text{e}^{-\icmpl p_0\tilde{t}}\;\mapsto\;-2\pi\icmpl\,\delta(p_0)\,\frac{\dint}{\dint p_0}\;.
\end{equation}
If $\Pi_{\text{R}}$ were an even function in $p_0$, the $p_0$-derivative evaluated at $p_0=0$ would simply vanish. Since all the $p_0$-dependence of $\Pi_{\text{R}}$ is carried by $W_\epsilon$, its pole structure generates a non-vanishing, but highly singular expression:
\begin{equation}
\label{DerivativeW}
  \left.\frac{\icmpl}{10}\frac{\dint}{\dint p_0} W_{\epsilon}(\omega_{12},p_0) \right|_{p_0=0} =-\frac{\pi}{5}\,\delta'(\omega_{12})\,.
\end{equation}
Accordingly, only the diagonal line, $\omega_{12}=\omega_1+\omega_2=0$, contributes to the shear viscosity. Denoting the perpendicular coordinate by $\omega=\frac 12(\omega_1-\omega_2)$, we obtain
\begin{equation}
 \label{etaZeroDerivativeW}
  \eta=\frac{4}{5\pi}\int\frac{\dint^3p}{(2\pi)^3}\; p^{(\mu}\,p^{\nu)} p_{(\mu}\,p_{\nu)} \int_{-\infty}^\infty\dint\omega\;F(\omega)\;,
\end{equation}
with the integrand:
\begin{figure}[b]
 \begin{center}
 \includegraphics[width=0.5\textwidth]{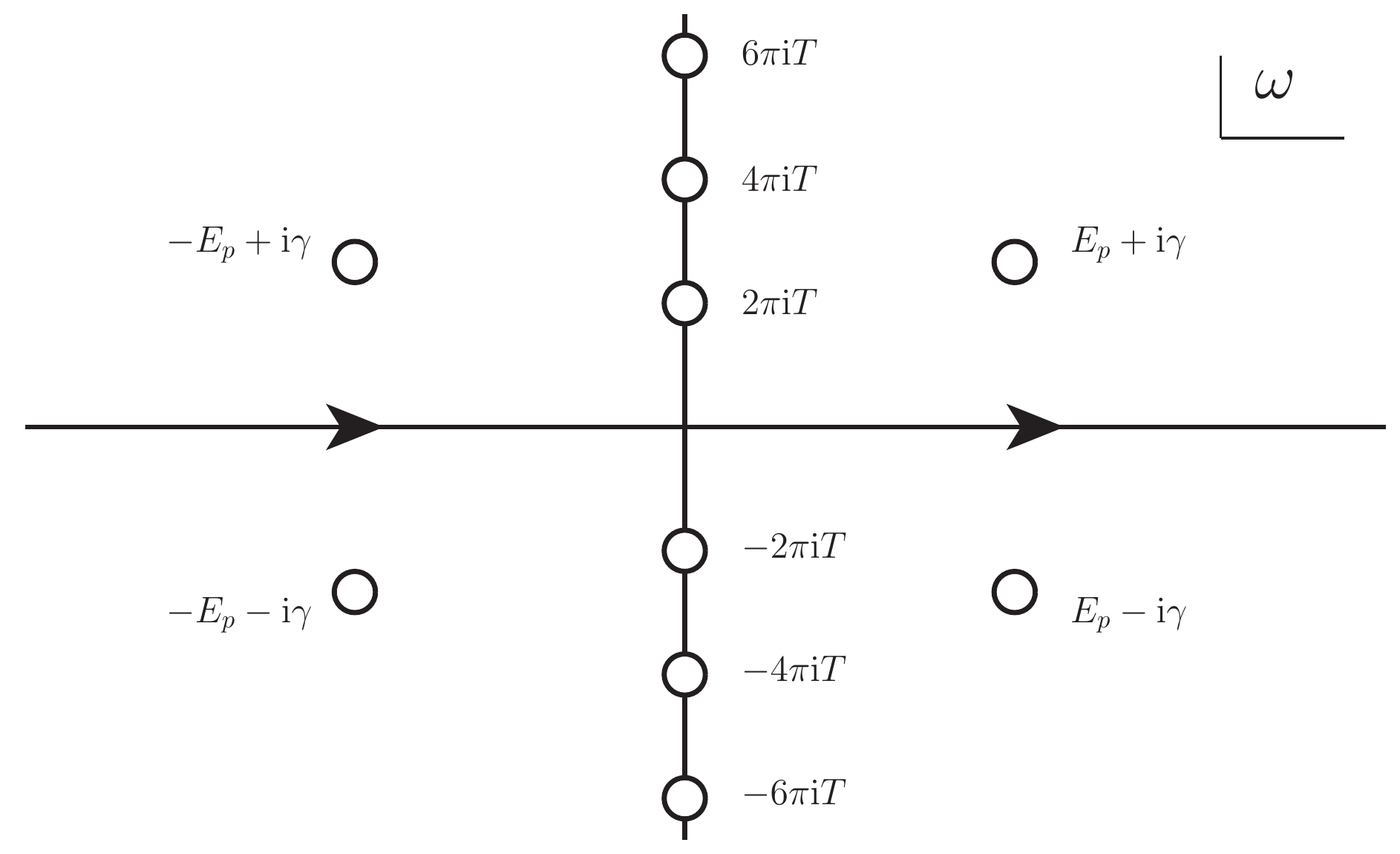}
  \end{center}
\caption{Pole structure of the integrand $F(\omega)$ in Eq.~\eqref{etaZeroDerivativeW}. For small $\gamma>0$ the $E_p$-dependent poles are close to the real axis. The Matsubara poles on the imaginary axis give no contribution to the shear viscosity at leading order in $\gamma$.}
 \label{FigurePoles4x2}
\end{figure}
\begin{equation}
  \label{etaZeroDerivativeWIntegrandF}
F(\omega)= \frac{2\,\omega^2 \text{e}^{\beta\omega}}{\left(\text{e}^{\beta\omega}-1\right)^2}\frac{\beta\gamma^2}{\left[E_p^2-(\omega-\icmpl\gamma)^2\right]^2 \left[E_p^2-(\omega+\icmpl\gamma)^2\right]^2}\;.
\end{equation}
There are infinitely many poles of $F(\omega)$ in the complex plane (Fig.~\ref{FigurePoles4x2}): the exponential function generates poles at $\omega=\icmpl\omega_n$, for $n\in\mathds Z\setminus\{0\}$, where $\omega_n=2\pi\icmpl nT$ are bosonic Matsubara frequencies. Indeed, the integrand $F(\omega)$ is regular at $\omega=0$. Furthermore, there are four poles (of second order): $\omega_j=\pm E_p\pm \icmpl\gamma$, for $j=1,2,3,4$. None of these poles is located on the real axis if $\gamma>0$. We use residue calculus to evaluate the integral, closing the contour in the upper half plane:
\begin{equation}
 \label{etaZeroDerivativeWResidua}
  \int_{-\infty}^\infty\dint\omega\; F(\omega)=2\pi\icmpl \left[\sum_{n=1}^\infty \text{Res}\, F(\icmpl\omega_n)+\sum_{i=1}^2 \text{Res}\,F(\omega_i)\right].
\end{equation}
The two residua at $\omega_{1,2}=\pm E_p+\icmpl\gamma$ can be calculated analytically (with some lengthly result) and their contribution to the shear viscosity can be expanded in a Laurent series in $\gamma$:
\begin{equation}
\label{etaZeroDerivativeWResiduaLaurent}
  2\pi\icmpl \sum_{i=1}^2 \text{Res}\,F(\omega_i) = \frac{\text{e}^{\beta E_p}}{\left(\text{e}^{\beta E_p}-1\right)^2}\frac{\pi\beta}{16 E_p^2\gamma}+\mathcal{O}(\gamma)\;.
\end{equation}
Note that there is no constant term in the expansion. The spectral width $\Gamma=2\gamma$ is assumed to be sufficiently small for this expansion to hold. In view of the perturbative origin of $\Gamma$ this assumption is justified.

Furthermore, one finds for the contribution of the $n$-th Matsubara frequency the following asymptotic behavior:
\begin{equation}
 \label{etaZeroDerivativeWResiduaMat}
  \text{Res}\,F(\icmpl\omega_n) \sim \frac{\beta^6\gamma^2}{n^9}\, ,\;\;\; \text{for small $\gamma,\beta$ and large $n\,$.}
\end{equation}
The sum of $\text{Res}\,F(\icmpl\omega_n)$ over $n\in\mathds N$ converges and yields a term of order $\mathcal{O}(\gamma^2)$ that can be neglected for small $\gamma$ compared to the dominant $\gamma^{-1}$ contribution in \eqref{etaZeroDerivativeWResiduaLaurent}.

Evaluating the integral \eqref{etaZeroDerivativeW} in the local rest frame to leading order in $\Gamma$, we arrive at the formula
\begin{equation}
 \label{etaZeroDelta2P4unev}
  \eta=
\frac{\beta}{30\pi^2}\int_0^\infty\dint p \,p^6\,\frac{n(E_p)[1+n(E_p)]}{E_p^2\,\Gamma(p)}+\mathcal{O}(\Gamma)\,.
\end{equation}
This is the final result for the shear viscosity $\eta$ in $g\phi^4$ theory. Note, that since $\Gamma(p)>0$, also $\eta>0$ follows, as expected from the second law of thermodynamics.

If one turns off the interaction and considers a free theory, the spectral width $\Gamma$ degenerates to a delta function, resulting in a divergent shear viscosity. This seemingly counter-intuitive behavior has been known for a long time \cite{TrainAnalogy}, not only for bosonic systems \cite{Fukutome08}.

\section{Shear Viscosity of a Hot Interacting Pion Gas}
\subsection{General Discussion}
Next we apply the skeleton expansion to pions within the framework of chiral perturbation theory ($\chi$PT). For $T<160\,\text{MeV}$ \cite{GasserLeut87, GerberLeut89, Aoki06, Bazavov09, Cheng10}, i.e.~in the temperature range where chiral symmetry is spontaneously broken, $\chi$PT can be applied with confidence. We use the second-order chiral Lagrangian, $\mathcal{L}_{2}$ (in $\sigma$-gauge), expanded up to fourth order in the pion fields:
\begin{equation}
 \label{ChPTLagrangian2pi4}
\begin{aligned}
\mathcal{L}_2 &=\frac  12\partial_\mu\vec{\pi}\cdot\partial^\mu\vec{\pi}-\frac{m_\pi^2}{2}\vec{\pi}\,^2 \\ 
& +\frac{1}{2f_\pi^2}\left(\vec{\pi}\cdot\partial_\mu\vec{\pi}\,\right)\left(\vec{\pi}\cdot\partial^\mu\vec{\pi}\,\right)-\frac{m_\pi^2}{8f_\pi^2}\left(\vec{\pi}\,^2\right)^2+\,\ldots
\end{aligned}
\end{equation}
In contrast to the discussion in $g\phi^4$ theory, not only the kinetic term, but also the momentum-dependent interaction terms contribute to the viscous-stress tensor \eqref{piLagrangePhi4Bare}:
\begin{equation}
 \label{ChPTpiTensor}
  \begin{aligned}
  \pi_{\mu\nu}=&(\Delta_\mu^\rho\Delta_\nu^\sigma-\frac 13\Delta_{\mu\nu}\Delta^{\rho\sigma})\\ &\cdot\Big[\left(\partial_\rho\vec{\pi}\,\right)\cdot\left(\partial_\sigma\vec{\pi}\,\right)+\frac{1}{f_\pi^2}\left(\vec{\pi}\cdot\partial_\rho\vec{\pi}\,\right)\left(\vec{\pi}\cdot\partial_\sigma\vec{\pi}\,\right)\Big].
  \end{aligned}
\end{equation}
The momentum-integrated thermal Green's function in Fourier space now has additional terms in comparison to $g\phi^4$ theory. The kinetic term is analogous to Eqs.~\eqref{spatIntThermGreen} and \eqref{correlation4point}:
\begin{equation}
  A=\frac{1}{V^2}\left\langle \mathcal{T}_\tau\left[\vec{\pi}(0)\cdot\vec{\pi}(0)\;\vec{\pi}(-\vec{p},\tau)\cdot\vec{\pi}(\vec{p},\tau)\right]\right\rangle_0\,.
\end{equation}
In addition there exist two terms at order $\mathcal{O}(f_\pi^{-2})$, and three terms at order $\mathcal{O}(f_\pi^{-4})$:
\begin{equation}
\label{feySkeletonBBC}
\begin{aligned}
  B_1&=\frac{1}{f_\pi^2V^4}\left\langle \mathcal{T}_\tau\left[\vec{\pi}^2(0)\, \vec{\pi}^2(-\vec{p},\tau)\, \vec{\pi}^2(\vec{p},\tau)\right]\right\rangle_0 \\ &=\parbox{0.15\linewidth}{\includegraphics[width=\linewidth]{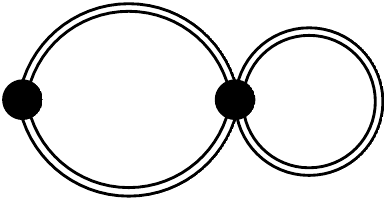}} + \;\parbox{0.2\linewidth}{\includegraphics[width=\linewidth]{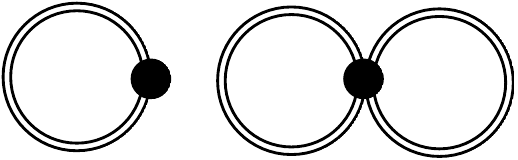}}\;,\\
  B_2&=\frac{1}{f_\pi^2V^4}\left\langle \mathcal{T}_\tau\left[\vec{\pi}^4(0)\, \vec{\pi}(-\vec{p},\tau)\cdot\pi(\vec{p},\tau)\right]\right\rangle_0 \\ &=\parbox{0.15\linewidth}{\includegraphics[width=\linewidth]{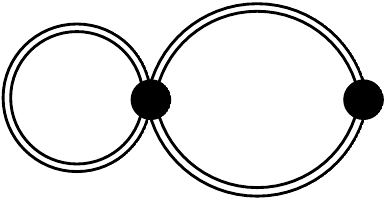}}+ \;\parbox{0.2\linewidth}{\includegraphics[width=\linewidth]{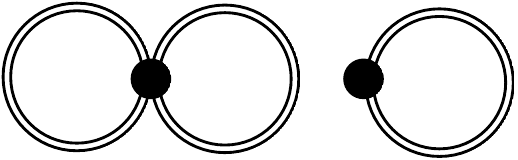}}\;,\\
  C&=\frac{1}{f_\pi^4V^6}\left\langle \mathcal{T}_\tau\left[\vec{\pi}^4(0)\, \vec{\pi}^2(-\vec{p},\tau)\vec{\pi}^2(\vec{p},\tau)\right]\right\rangle_0 \\
  &=\;\parbox{0.1\linewidth}{\includegraphics[width=\linewidth]{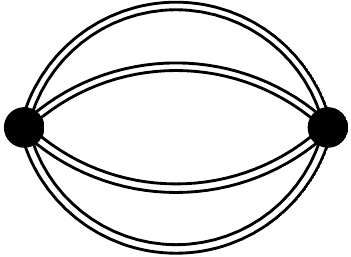}} + \;\parbox{0.2\linewidth}{\includegraphics[width=\linewidth]{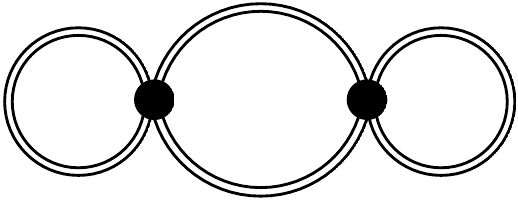}}+\;\parbox{0.2\linewidth}{\includegraphics[width=\linewidth]{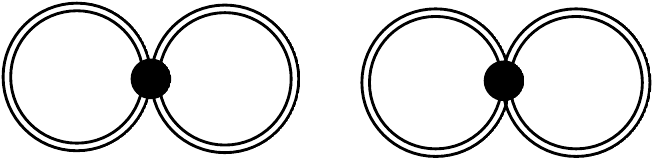}}\;.
\end{aligned}
\end{equation}
The disconnected parts of the skeleton expansion are vacuum loops and can therefore be dropped. Note again that the diagrammatic representation already uses $G_\beta(-\vec{p},\tau)=G_\beta(\vec{p},\tau)$. Furthermore, the first terms of $B_1,B_2$ and the second one of $C$ can be absorbed by the skeleton vertex (or eliminated by a field renormalization), since we are anyhow expanding in \textit{fully-dressed} quantities. At order $\mathcal{O}(f_\pi^{-4})$ there is an additional term in the skeleton expansion giving rise to a correction of the momentum-integrated thermal Green's function:
\begin{equation}
 \label{LOskeletonTemporalChPT}
  \begin{aligned}
  \Pi_\beta(\omega_n)&=2\left(N_\text{f}^2-1\right)\int_{0}^\beta \dint\tau\; \text{e}^{\icmpl\omega_n\tau}\int \frac{\dint^3p}{(2\pi)^3} \\ 
  & \hspace{-0.8cm}\cdot p^{(\mu}\,p^{\nu)} p_{(\mu}\,p_{\nu)}\left[G^2_\beta(\vec{p},\tau)+\frac{12}{f_\pi^4V^2}G^4_\beta(\vec{p},\tau)\right].
  \end{aligned}
\end{equation}
Additonally to the calculation in $g\phi^4$ theory \eqref{LOskeletonTemporal} we have now the flavor prefactor $N_\text{f}^2-1=3$ which arises from the sum over the isospin degree of freedom. The leading order term in Eq.~\eqref{LOskeletonTemporalChPT} has the symmetry factor $2$, whereas the correction term has the combinatorial factor $4!=24\,$.

We have started with the second-order chiral Lagrangian and arrived at the skeleton-expanded thermal Green's function $\Pi_\beta(\omega_n)$ including corrections of order $\mathcal{O}(f_\pi^{-4})$. In the thermodynamic limit, $V\to\infty$, this correction term vanishes. Thus we find for the shear viscosity in $\chi$PT the same functional formula $\eta[\Gamma]$ as in $g\phi^4$ theory, Eq.~\eqref{etaZeroDelta2P4unev}, but corrected by an isospin factor three:
\begin{equation}
 \label{directCompPhiFourChPT}
  \eta|_{\chi\text{PT}}=
\frac{\beta}{10\pi^2}\int_0^\infty\dint p \,p^6\,\frac{n(E_p)[1+n(E_p)]}{E_p^2\,\Gamma(p)}\,,
 \end{equation}
with $E_p^2=\vec{p}\,^2+m_\pi^2$. We mention that this identity does also hold if one allows for higher-order terms in the chiral expansion or higher-order terms in the field expansion of $\mathcal{L}_2$. All these corrections would be of higher order in the coupling $1/f_\pi$ and suppressed by higher powers of the inverse volume.

\subsection{Analytic Evaluation of the Spectral Width}
\begin{figure}[b]
\begin{center}
  \includegraphics[width=0.35\textwidth]{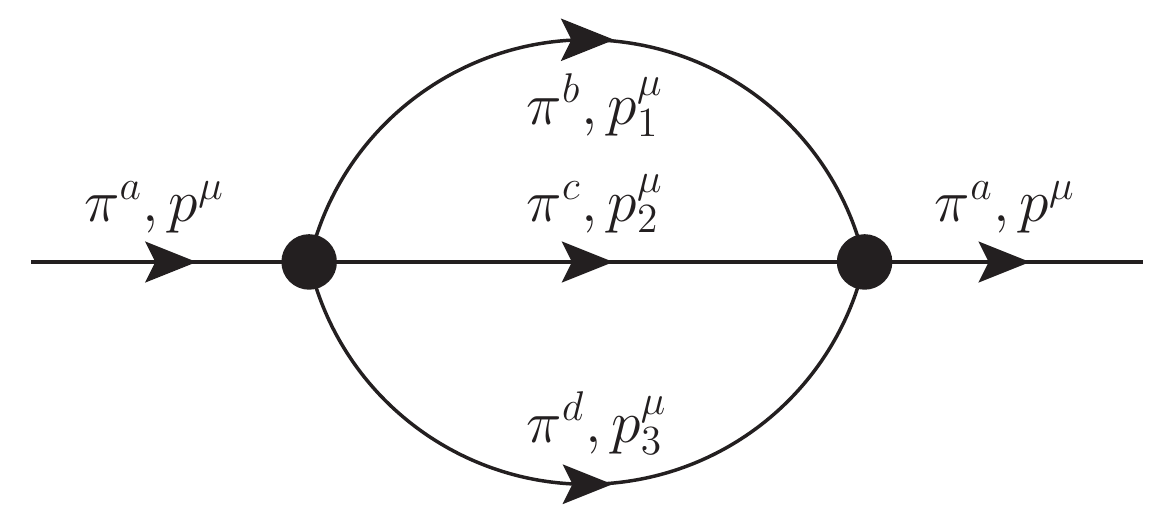}
\end{center}
\caption{Self-energy diagram $\Sigma_\beta^{(2)}$ of the pion in the heat bath. The arrows indicate the direction of the three-momenta.}
\label{figChPT12selfenergy}
\end{figure}

Once the spectral width $\Gamma$ is known, the shear viscosity $\eta$ in Eq.~\eqref{directCompPhiFourChPT} can be computed. In $\chi$PT the leading-order contribution to $\Gamma$ comes from the two-loop diagram shown in Fig.~\ref{figChPT12selfenergy}. The self-energy diagram at one-loop level is real, hence it gives no contribution to the spectral width.

\begin{figure}[b]
\begin{tabular}{cc}
  \includegraphics[width=0.18\textwidth]{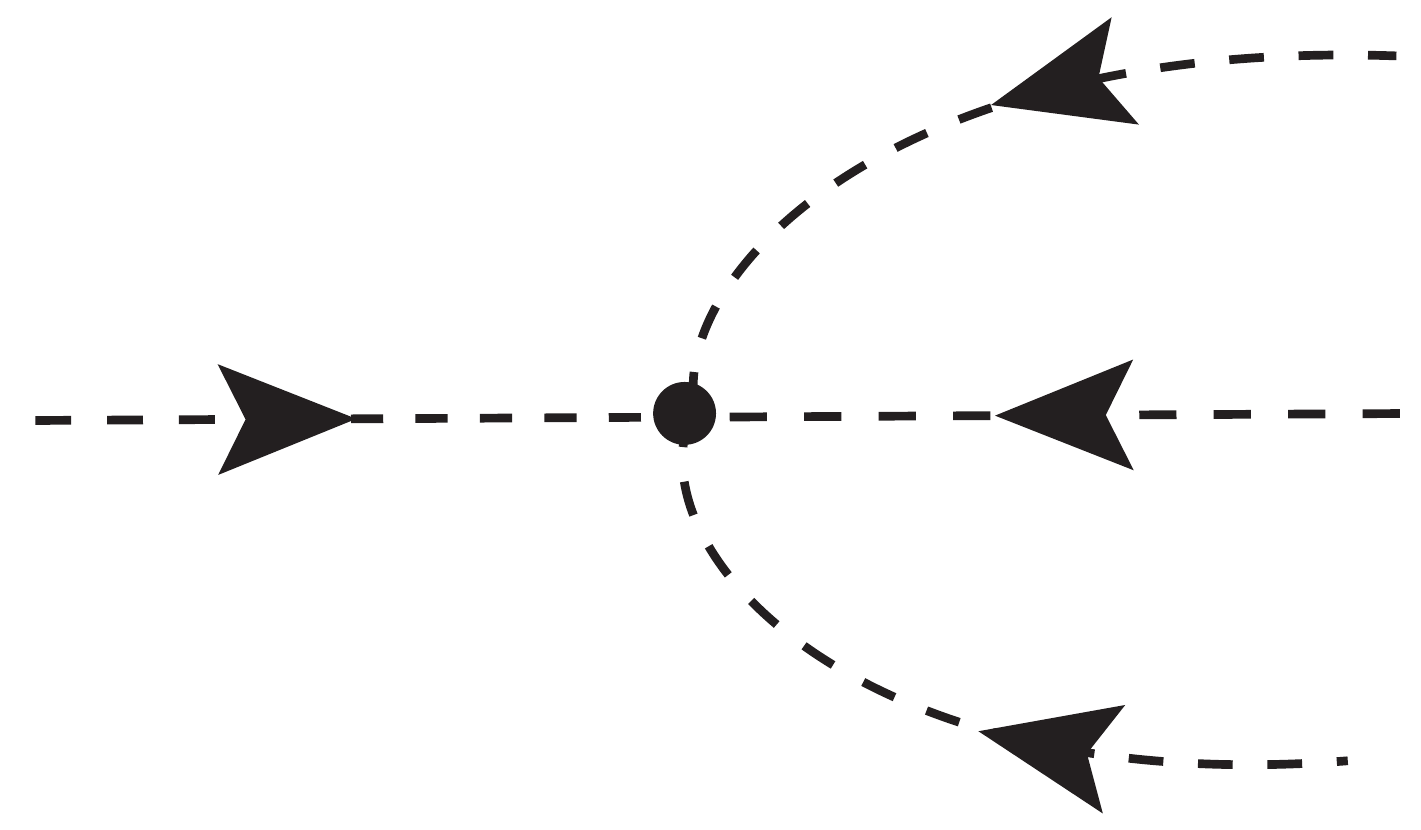}  &  \hspace{1cm} \includegraphics[width=0.18\textwidth]{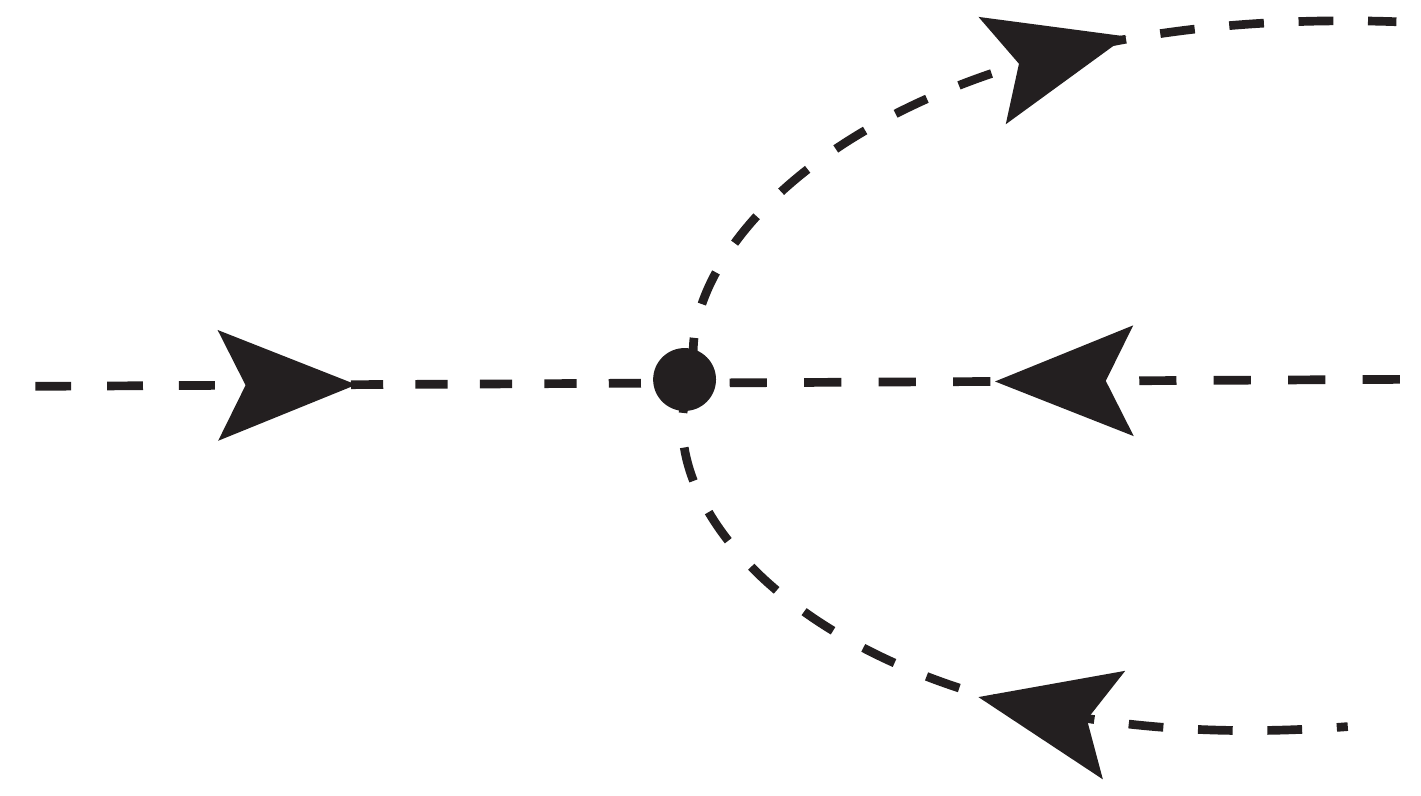}  \\
  $(+++)$ & \hspace{1cm} $(-++)$ \\
  \includegraphics[width=0.18\textwidth]{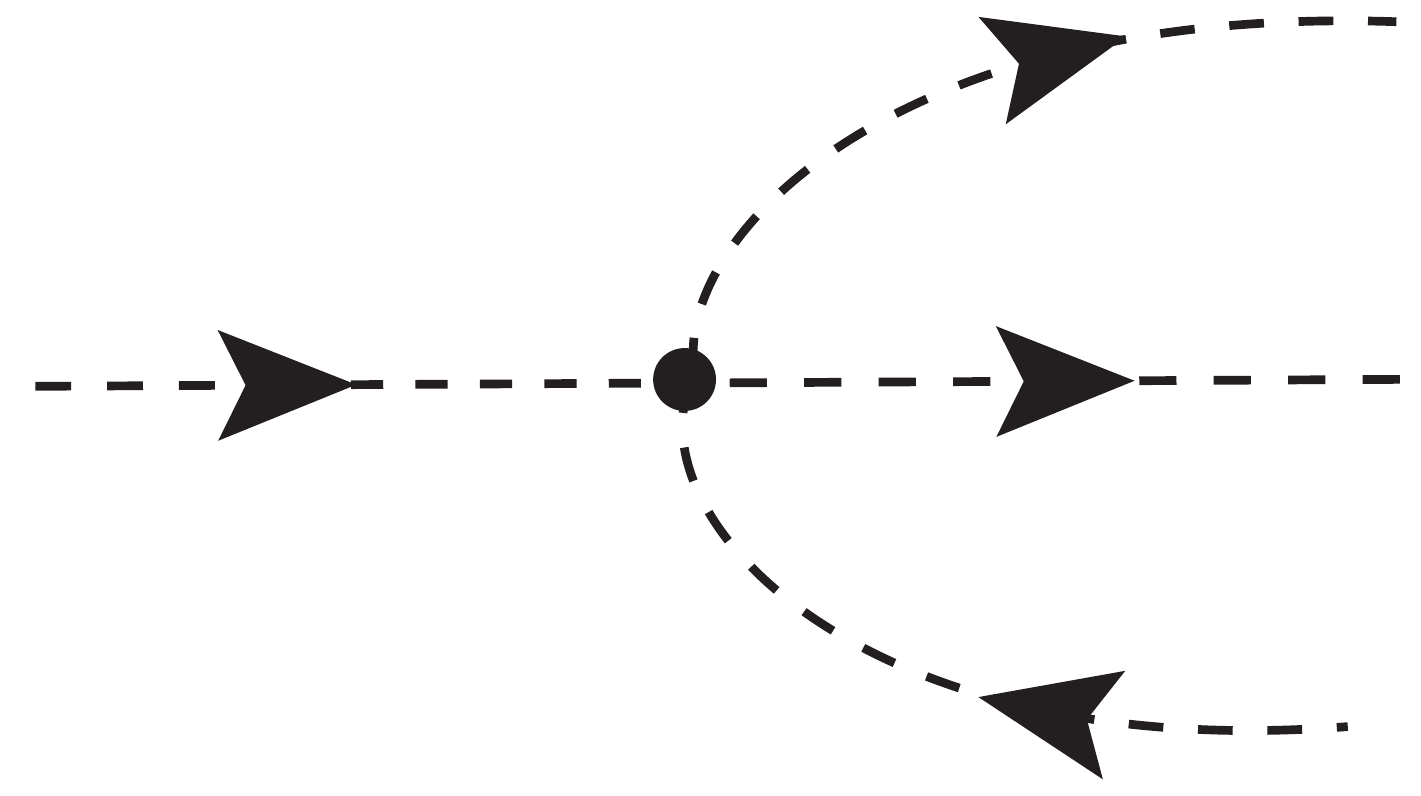}  &\hspace{1cm} \includegraphics[width=0.18\textwidth]{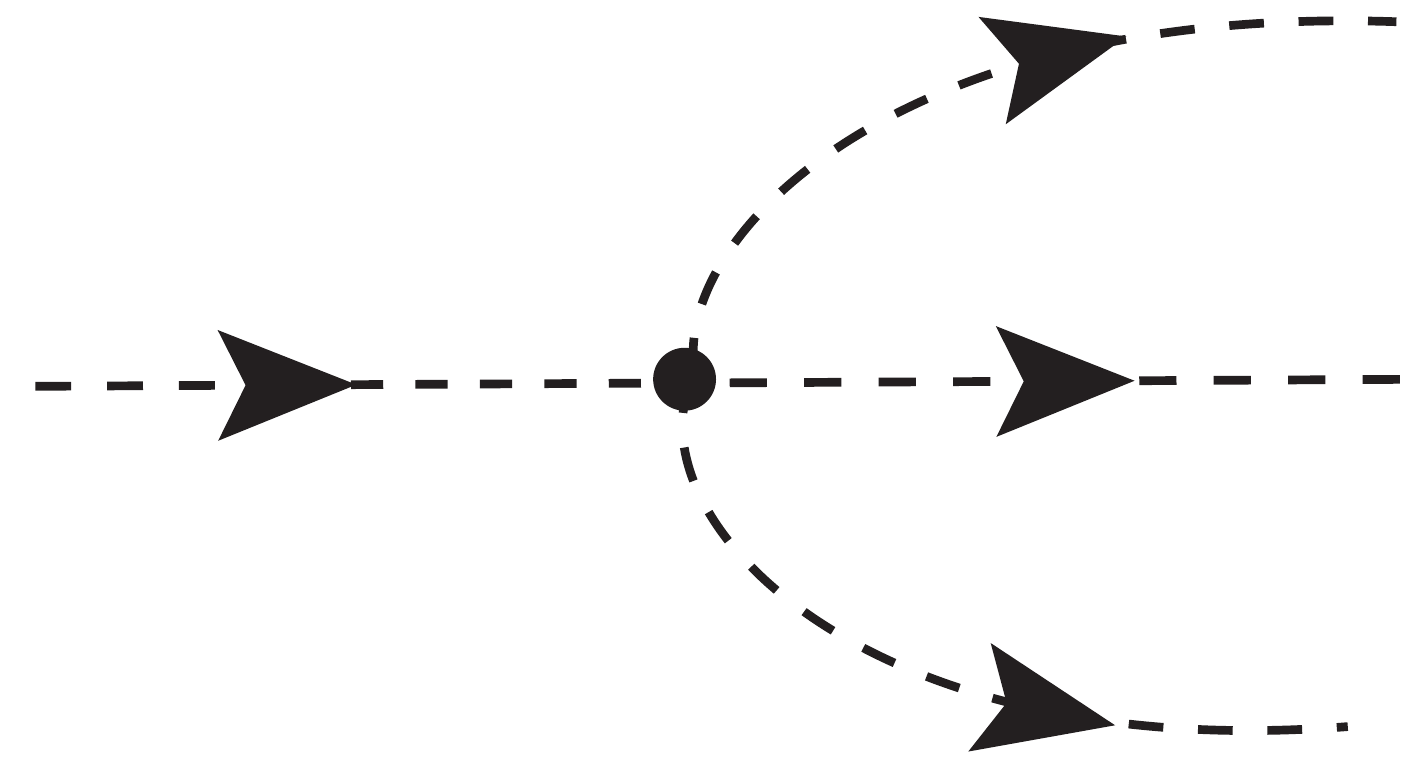}  \\ 
  $(--+)$ & \hspace{1cm} $(---)$
\end{tabular}
\caption{Relative signs of energies in the two-loop diagram with respect to the incoming-particle energy $p_0$. Only the $2\to 2$ scattering process with signature $(--+)$ gives an on-shell contribution to the spectral width $\Gamma(p)$ in Eq.~\eqref{SpectralWidthChPTFinal}. The dashed lines symbolize the on-shell condition at the vertices.}
\label{SignsPartialDecompFigure}
\end{figure}

In the two-loop diagram in Fig.~\ref{figChPT12selfenergy} the two Matsubara sums over $(m_1,m_2)\in\mathds{Z}^2$, can be performed analytically. Taking into account the energy-momentum conservation, $E_3^2=(\vec{p}-\vec{p}_1-\vec{p}_2)^2+m_\pi^2$, and the on-shell identity for the Mandelstam variables, $s+t+u=4m_\pi^2$, we arrive at an expression for the thermal self-energy $\Sigma_\beta(\icmpl\omega_n,\vec{p}\,)$ which can be analytically continued via $\icmpl\omega_n\mapsto p_0+\icmpl\epsilon$ to the retarded self-energy $\Sigma_{\text{R}}$. The spectral width of a pion with four-momentum $p^\mu=(p_0,\vec{p}\,)$ reads:

\begin{widetext}
\begin{equation}
\label{SpectralWidthChPTFinal}
  \Gamma(p)=\frac{1}{4f_\pi^4}\frac{1}{n(E_p)E_p} \left(\prod_{i=1}^3\int\frac{\dint^3p_i}{(2\pi)^3\,2E_i}\right) \big[2(s^2+t^2+u^2)-9m_\pi^4\big] n(E_1)n(E_2)[1+n(E_3)] (2\pi)^4\delta^{(4)}(p\!-\!p_1\!-\!p_2\!+\!p_3)\;.
\end{equation}
\end{widetext}
The combination of relative signs in the delta function $\delta^{(4)}(p-p_1-p_2+p_3)$ is the only one that gives rise to an on-shell contribution to the imaginary part $\text{Im}\,\Sigma_{\text{R}}$. In all other cases the energy-conservation delta function leads to zero. In general, i.e. in the off-shell case, there are four different combinations of relative signs, shown in Fig.~\ref{SignsPartialDecompFigure}.
The first case leads to $\delta(p_0+E_1+E_2+E_3)=0$, hence this realization gives no contribution. The structures $(-++)$ and $(---)$ are not realized on-shell, because they would describe the decay of a particle into three copies of it. This is not possible for massive pions. Thus, only three equal terms with $(--+)$ signature remain.

From a numerical point of view the representation of $\Gamma$ in Eq.~\eqref{SpectralWidthChPTFinal} is only symbolic. The (four-dimensional) delta function and the nine-dimensional integral over it are not well-conditioned for numerical evaluation. As shown in the appendix \ref{AppCalc}, one can rewrite the expression for the spectral width as a two-dimensional energy integral suitable for numerical evaluation. This procedure introduces the quantities (here $p_i=|\vec{p}_i|$)
\begin{equation}
\begin{aligned}
Q_-&=\text{max}\left\{|p_2-p_3|,|p-p_1|\right\}\,,\\ Q_+&=\text{min}\left\{p_2+p_3,p+p_1\right\}\,.
\end{aligned}
\end{equation}
The integrand in Eq.~\eqref{GammaXPTNumericalSinh} depends on differences of powers of $Q_\pm$,
\begin{equation}
  X(Q_-,Q_+)=\sum_{i\in\{0,\pm 2,\pm 4\}} X_i\left(Q_+^{i+1}-Q_-^{i+1}\right),
\end{equation}
where the five coefficient functions $X_i$ have the mass dimension $\text{dim}\, X_i=4-i$ and depend on the pion mass $m_\pi$ and on the energies $E_2, E_3, E_p$ only. Their explicit forms are given in the appendix, \eqref{GammaXPTNumericalXCoeffP4} to \eqref{GammaXPTNumericalXCoeffM4}. In the end the double-integral representation of the spectral width reads
\begin{equation}
\label{GammaXPTNumericalSinh}
\begin{aligned}
  \Gamma(p)= &\frac{\sinh\frac{\beta E_p}{2}}{2pE_pf_\pi^4\;(8\pi)^3}\int_{m_\pi}^\infty\dint E_3 \int_{m_\pi}^{E_p+E_3-m_\pi}\dint E_2\\ 
  &\hspace{0.2cm}\cdot\frac{X(Q_+,Q_-)\,\Theta(Q_+-Q_-)}{\sinh\frac{\beta E_1}{2}\,\sinh\frac{\beta E_2}{2}\,\sinh\frac{\beta E_3}{2}}\;.
\end{aligned}
\end{equation}
In comparison with Ref. \cite{GoityLeut89}, where the mean free path of hot pions has been calculated in the infinite-volume limit (consistent with our skeleton-expansion approach), we have found an analytical representation of the spectral width with a different algebraic structure. Our result involves the rational function $X(Q_+,Q_-)$ whereas a polynomial function of the energies appears in Ref. \cite{GoityLeut89}. In addition, the max/min pattern in the two representations are quite different and it is not apparent how to convert one into another.

Irrespective of that our results for the spectral width are identical with those of Ref.~\cite{GoityLeut89}: only the integrands differ, but the numerical values of the integrals for the mean free path coincide. In essence, the mean free path $\lambda(p)$ is the inverse spectral width:
\begin{equation}
\label{DefMeanFreePath}
  \lambda(p)=\frac{p}{E_p\Gamma(p)}\mathop{\xrightarrow{\hspace*{1cm}}}_{m_\pi\to 0} \;\frac{1}{\Gamma(p)}\;.
\end{equation}
As an aside, we mention that the spectral width $\Gamma(p)$ in $g\phi^4$ theory can be obtained by inserting the expression $X\!=\!576f_\pi^4g^2(Q_+-Q_-)$ into Eq.~\eqref{GammaXPTNumericalSinh} and replacing $m_\pi$ by $m$. This result for $X$ is derived from the two-loop diagram in Fig.~\ref{figChPT12selfenergy} as well, but evaluated in $g\phi^4$ theory.
\begin{figure}[t]
  \includegraphics[width=0.49\textwidth]{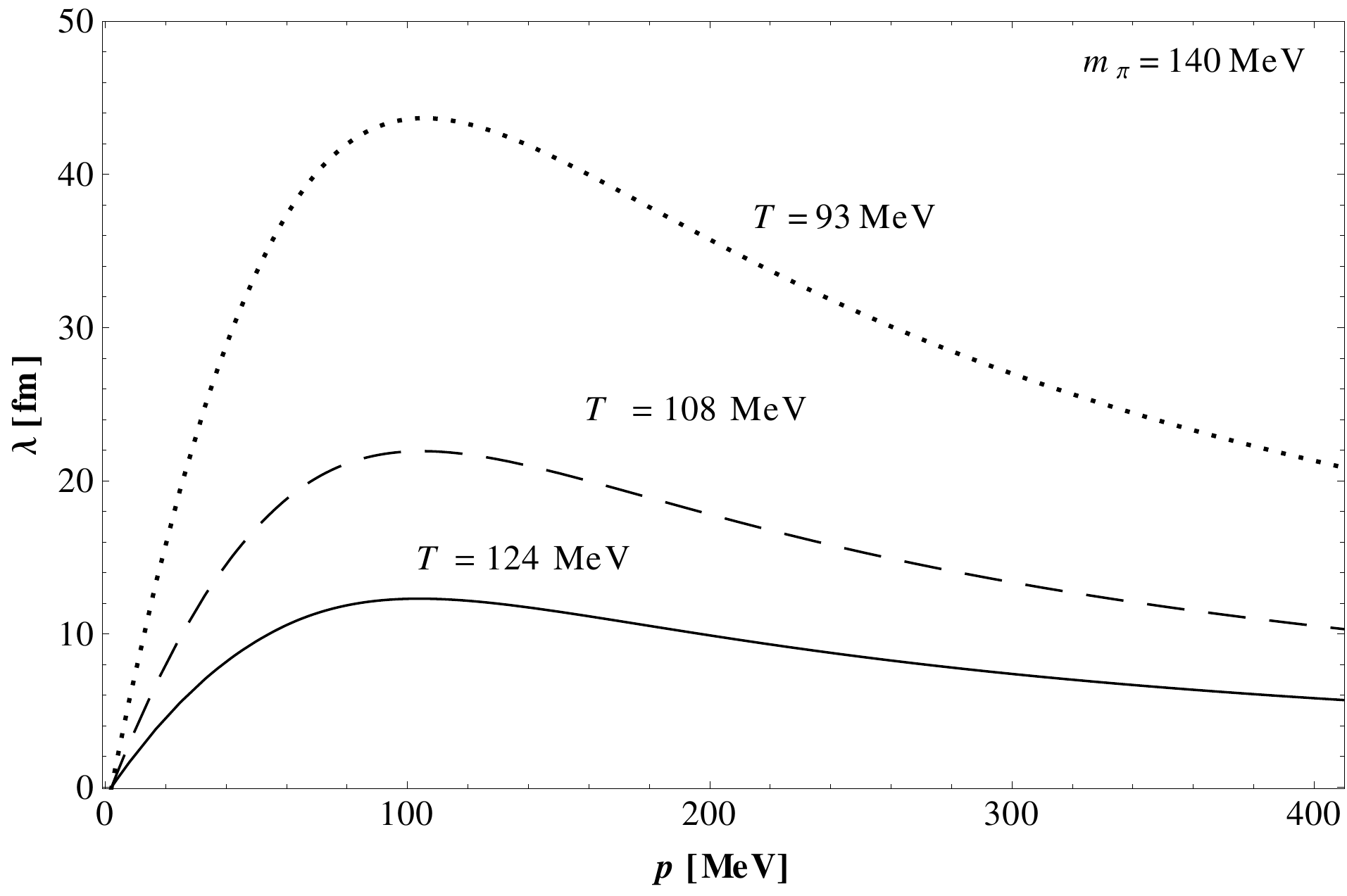} \\
  \includegraphics[width=0.5\textwidth]{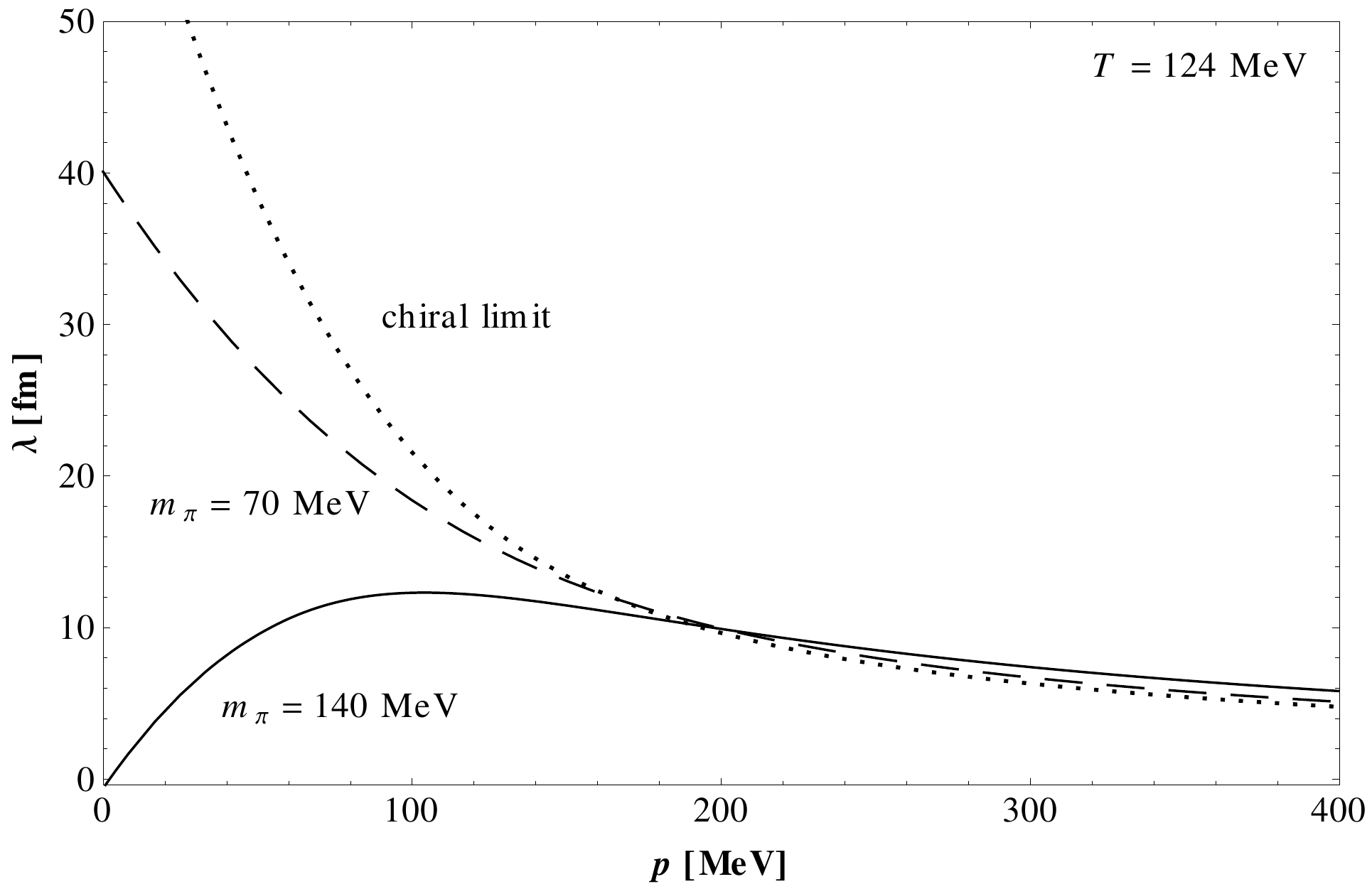}
\caption{Mean free path $\lambda(p)$ of a hot pions for different temperatures (upper figure) and pion masses (lower figure). The two solid lines in the figures are drawn for the same combination of mass and temperature: $(m_\pi,T)=(140\,\text{MeV},124\,\text{MeV})$.}
\label{FigureMeanFreePath}
\end{figure}
 
\subsection{Numerical Results for \boldmath{$\lambda$} and \boldmath{$\eta$}}
The numerical results for the mean free path $\lambda(p)$ are shown for different temperatures and pion masses in Fig.~\ref{FigureMeanFreePath}. For increasing temperatures the abundance of thermal pions lowers the mean free path, but the maximum position $p\approx 100\,\text{MeV}$ is almost independent of the temperature. The mass dependence of $\lambda(p)$ is more pronounced: when switching from the physical value, $m_\pi=140\,\text{MeV}$, to pions with half the physical mass and to the chiral limit, the mean free path features a monotonic increase at low momenta $p\lesssim 200\,\text{MeV}$. For all temperatures and masses the mean free path decreases for large momenta. In the numerical evaluation we have used the value $f_\pi=93\,\text{MeV}$ for the pion decay constant.

\begin{figure}[t]
  \begin{center}
 \includegraphics[width=0.5\textwidth]{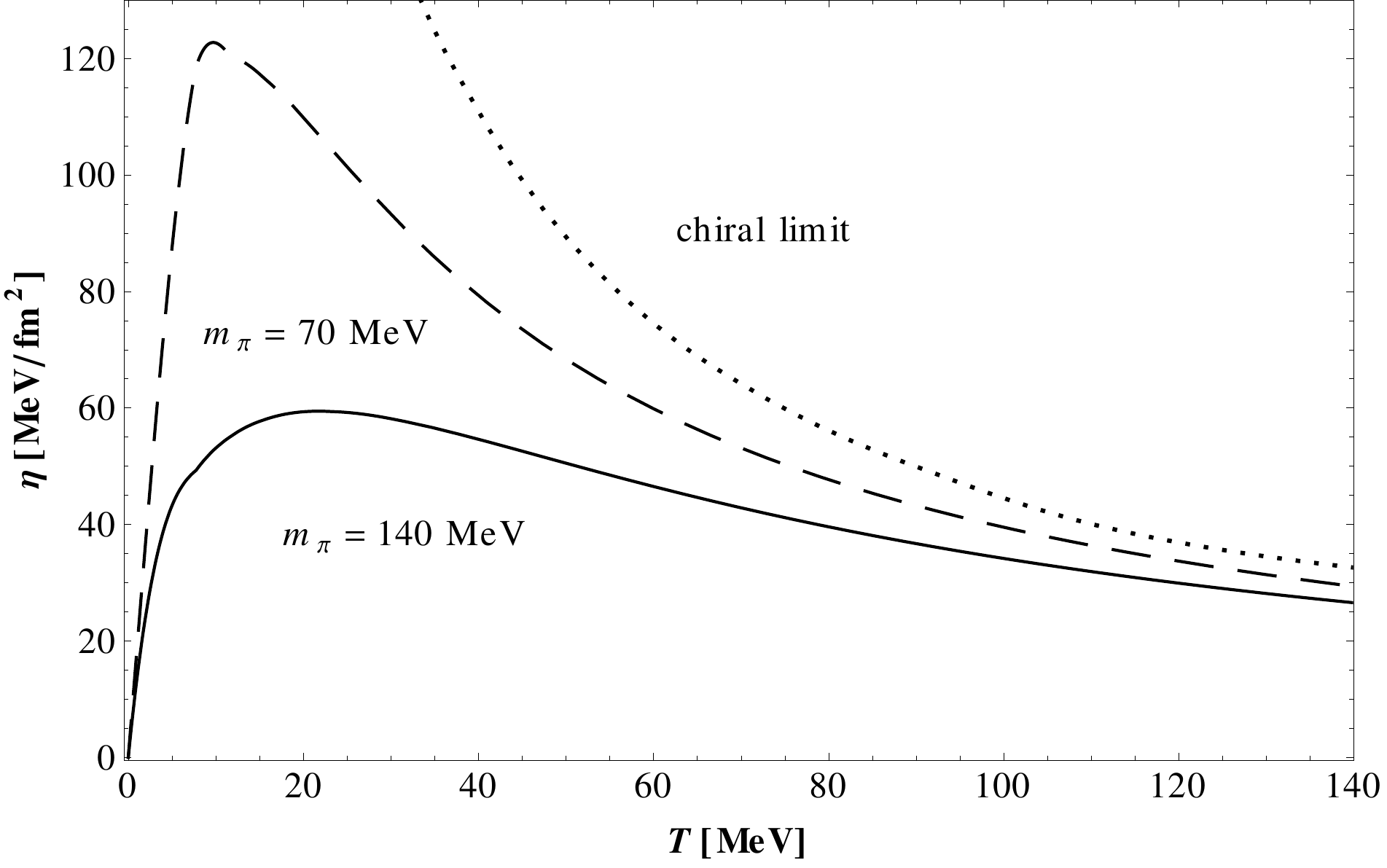}
    \end{center}
\caption{Shear viscosity $\eta(T)$ of an interacting pion gas for different pion masses. In the chiral limit one gets $\eta\sim 1/T$ as expected from dimensional analysis.}
\label{FigureXPTEtaMasses}
\end{figure}

Interpolating the numerical results for the mean free path at different temperatures and masses finally allows to calculate the shear viscosity. The results are shown in Fig.~\ref{FigureXPTEtaMasses}. In the chiral limit, $m_\pi=0$, one finds a reciprocal dependence on the temperature, $\eta(T)\sim f_\pi^4/T$, as expected from dimensional analysis. For the physical pion mass, $m_\pi=140\,\text{MeV}$, the maximum of $\eta$ is located at $T\approx20\,\text{MeV}$. With decreasing pion masses, this maximum moves to lower temperatures. For high temperatures the shear viscosity depends only weakly on the pion mass.

Note, that the function $\eta(T;m_\pi)$ is not continuous at the origin, since
\begin{equation}
 \label{XPTLimitsMandT}
   0=\lim_{m_\pi\to 0}\;\underbrace{\lim_{T\to 0}\eta(T;m_\pi)}_{=0}\neq \lim_{T\to 0}\;\underbrace{\lim_{m_\pi\to 0}\eta(T;m_\pi)}_{\sim 1/T} =\infty\,.
\end{equation}
In Fig.~\ref{FigureXPTEtaMasses} we have chosen the units \text{MeV/fm}$^2$ instead of \text{MeV}$^3$ in order to meet the classical interpretation of shear viscosity.

\section{Ratio \boldmath{$\eta/\lowercase{s}$} for the Pion Gas}
In 1998 Maldacena \cite{Maldacena99} demonstrated that, under certain conditions, there is a duality between superstring theory and superconformal field theory. More precisely, the AdS/CFT correspondence between supergravity on five-dimensional anti-de Sitter space, $\text{AdS}_5$, and four-dimensional $\mathcal{N}=4$ superconformal ${\rm SU}(N_\text{c})$ Yang-Mills theory in the 't~Hooft limit has been proven. In this special case the ratio $\eta/s$ (shear viscosity over entropy density) is equal to the KSS lower bound $1/4\pi$ \cite{Kovtun05}. In 2005 the KSS conjecture has been formulated:
\begin{quote}
Most quantum field theories do not have simple gravity duals. ($\ldots$) We speculate that the ratio $\eta/s$ has a lower bound $\eta/s\geq 1/4\pi$ for all relativistic quantum field theories at finite temperature and zero chemical potential.
\end{quote}
So far, the KSS lower bound for $\eta/s$ is respected by experimental results for a wide variety of thermal systems \cite{Lacey07}. The fundamental theory describing heavy-ion collisions is QCD which does not possess a gravity dual: QCD is neither supersymmetric nor conformal. The classical scale independence of the QCD Lagrangian is broken anomalously by quantum effects resulting in a running coupling $\alpha_\text{s}(q^2)$. In addition, QCD is a ${\rm SU}(3)$ gauge theory and can be described by the large-$N_\text{c}$ limit only approximatively. Since 2007 counterexamples to the KSS conjecture have been established \cite{Cohen07,Rebhan12}.
\begin{figure}[t]
  \begin{center}
 \includegraphics[width=0.5\textwidth]{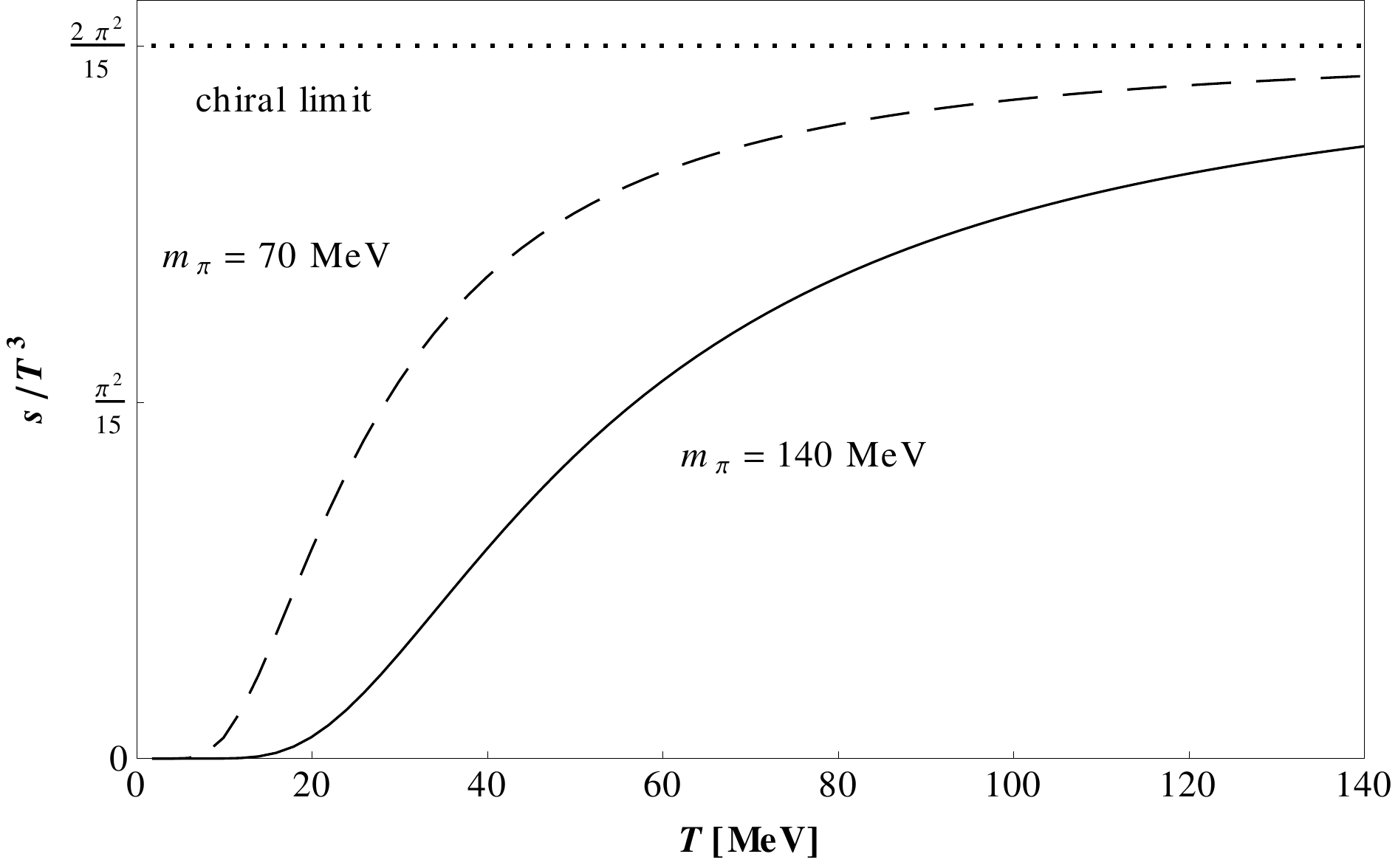}
    \end{center}
\caption{Entropy density $s(T)$ of an interacting pion gas for different pion masses. We compare the $\chi$PT results to the Stefan-Boltzmann limit $s(T)=\frac{2\pi^2}{15}T^3$ of three massless non-interacting pions.}
 \label{FigureEntropyDensityChPT}
\end{figure}

The temperature-dependent entropy density of the interacting pion gas reads at two-loop order \cite{GerberLeut89}
\begin{equation}
\label{EntropyDensityChPT}
\begin{aligned}
  s(T) &=\frac{T}{2\pi^2}\left[4T^2h_5(\beta m_\pi)+3m_\pi^2h_3(\beta m_\pi)\right] \\
   &-\frac{3m_\pi^2T}{16\pi^4f_\pi^2}\,h_3(\beta m_\pi)\left[2T^2h_3(\beta m_\pi)+m_\pi^2h_1(\beta m_\pi)\right],
\end{aligned}
\end{equation}
with the integral function for $\xi\geq 0$\,:
\begin{equation}
  h_n(\xi)=\int_\xi^\infty\dint x\,\frac{(x^2-\xi^2)^{\frac{n}{2}-1}}{\text{e}^x-1}\;.
\end{equation}
In the limit $f_\pi\to\infty$ only the first line of Eq.~\eqref{EntropyDensityChPT} contributes to the entropy density. The correction to $s$ at order $\mathcal{O}(f_\pi^{-2})$ is less than $1.5\%$ (for $m_\pi=140\,\text{MeV}$) and decreases even more for lower masses. In the chiral limit the entropy density is given by
\begin{equation}
\label{EntropyDensityChiralLimit}
  \lim_{m_\pi\to 0} s(T)=\frac{2\pi^2}{15}T^3\;.
\end{equation}
This is the well-known Stefan-Boltzmann limit for three non-interacting massless pions. In Fig.~\ref{FigureEntropyDensityChPT} we compare the reduced entropy density $s/T^3$ of the pion gas in $\chi$PT to the constant $2\pi^2/15$ from Eq.~\eqref{EntropyDensityChiralLimit}.

\begin{figure}[t]
  \begin{center}
 \includegraphics[width=0.5\textwidth]{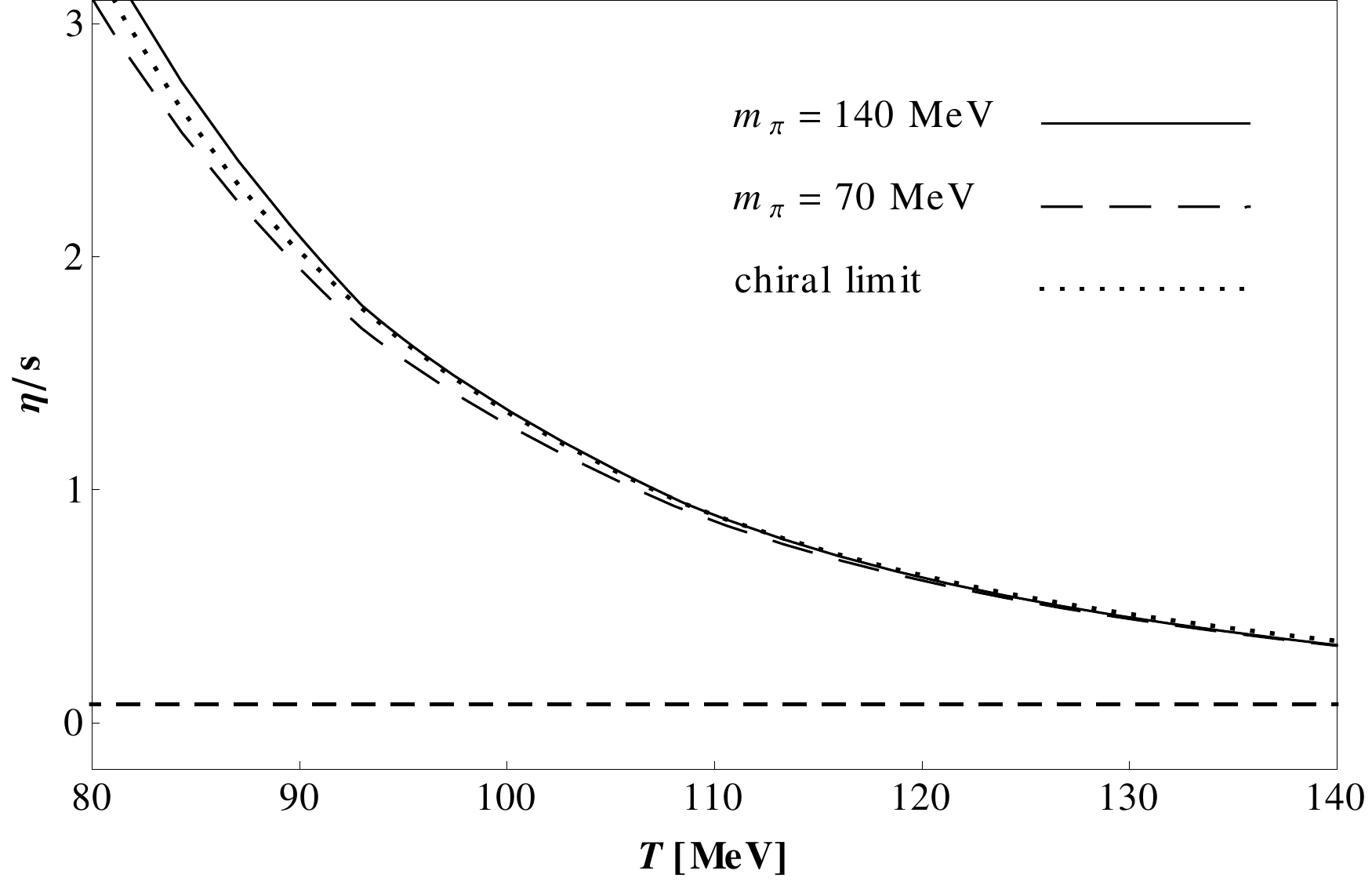}
    \end{center}
\caption{Ratio between shear viscosity and entropy density $\eta(T)/s(T)$ of an interacting pion gas for different pion masses. The horizontal dashed line is the lower bound $1/4\pi$ from AdS/CFT correspondence.}
 \label{FigureEtasRatio}
\end{figure}

In Fig.~\ref{FigureEtasRatio} we show the temperature dependence of the ratio $\eta/s$ for the interacting pion gas. Actually, for temperatures up to $T\approx 140\,\text{MeV}$ we find a ratio $\eta/s$ which is well above the KSS lower bound $1/4\pi$. In the considered temperature range $80\,\text{MeV}\!<\!T\!<\!140\,\text{MeV}$ the ratio $\eta/s$ is almost independent of the pion mass. Our results compare qualitatively well with alternative approaches to calculate the shear viscosity of the pion gas \cite{NicolaGomez06,NicolaGomez08}.

\section{Summary}
In this work we have calculated, on the basis of the Kubo-type formula \eqref{ShearViscosity}, the shear viscosity $\eta$ of a pion gas within chiral perturbation theory ($\chi$PT). The skeleton expansion, corresponding to an expansion in full propagators, has been applied to the thermal (four-point) Green's function. At one-loop level this expansion leads to the squared Matsubara propagator (two-point function). This technique has first been explored in $g\phi^4$ theory, and then applied to $\chi$PT.
We have found that at one-loop level of the skeleton expansion and in the thermodynamic limit $V\!\to\!\infty$, the shear viscosities for real scalar fields, Eq.~\eqref{etaZeroDelta2P4unev}, and pseudoscalar pions, Eq.~\eqref{directCompPhiFourChPT}, differ only by an isospin factor. The functional dependence of $\eta$ on the spectral width coincides for both cases.

The spectral width $\Gamma(p)$ of pions has been calculated to two-loop order in thermal $\chi$PT resulting in a new analytical representation Eq.~\eqref{GammaXPTNumericalSinh}. We have found that the shear viscosity of the pion gas reaches a maximum at $T\approx 20\,\text{MeV}$ in the case of physical pion mass $m_\pi\!=\! 140\,\text{MeV}$. Furthermore we have investigated the ratio $\eta/s$ (with $s$ the entropy density) and found that it decreases monotonously with rising temperature but exceeds the KSS lower bound $1/4\pi$ for all temperatures $T<140\,\text{MeV}$ where $\chi$PT is applicable.

\appendix
\section{Details on the Analytical Evaluation of the Spectral Width} \label{AppCalc}
We outline the analytical calculation which leads to the numerically well-conditioned expression \eqref{GammaXPTNumericalSinh} for the spectral width $\Gamma(p)$. The momentum integral $\int \dint^3p_1$ in Eq.~\eqref{SpectralWidthChPTFinal} is canceled by the three-dimensional delta function, hence we can express the corresponding energy $E_1$ by the incoming momentum and the two remaining internal momenta:
\begin{equation}
 \label{Energy1EtaXPT}
  \begin{aligned}
   E_1^2=E_p^2+|\vec{p}_3-\vec{p}_2|^2+2p|\vec{p}_3-\vec{p}_2|\cos\theta_p\;.
\end{aligned}
\end{equation}
We integrate out the dependences on $\cos\theta_p$ and $\varphi_p$, using
\begin{equation}
 \label{EnergyEtaXPTformalIntegration}
  1=\frac{1}{2\pi}\int_0^{2\pi}\dint\varphi_p\;\frac{1}{2}\int_{-1}^1\dint\cos \theta_p\, .
\end{equation}
Denoting the squared on-shell pion-pion scattering amplitude by $V_{\text{on}}^2=[2(s^2+t^2+u^2)-9m_\pi^4]/f_\pi^4$ and performing the $\cos\theta_p$ integral, the energy-conserving delta function fixes the value of $\cos\theta_p$:
\begin{equation}
 \label{GammaXPTNumericalDeltaE}
   \begin{aligned}
    &\frac 12\int_{-1}^1\dint\cos\theta_p\, \frac{V_{\text{on}}^2(\cos\varphi_p,\cos\theta_p)}{2E_1}\,\delta(E_p\!-\!E_1\!-\!E_2\!+\!E_3)\\
    &\hspace{2cm}=\frac{1}{4pQ}V_\text{on}^2(\cos\varphi_p,x_0)\,,
   \end{aligned}
\end{equation}
where we have introduced the abbreviations $Q=|\vec{p}_3-\vec{p}_2|$ and $x_0=\frac{1}{2pQ}\left[(E_3-E_2)(E_3-E_2+2E_p)-Q^2\right]$. In addition to $\theta_p=\sphericalangle(\vec{p},\vec{p}_3-\vec{p}_2)$ and $\varphi_p$\,, we introduce the following quantities:
\begin{equation}
 \label{GammaXPTNumericalDefAnglesCo}
   \begin{aligned}
   \varphi &=\sphericalangle(\vec{p},\vec{p}_3+\vec{p}_2)\, , \\
   \Delta &=\sphericalangle(\vec{p}_3-\vec{p}_2,\vec{p}_3+\vec{p}_2)\, , \\
   R &=|\vec{p}_3+\vec{p}_2|\, .
  \end{aligned}
\end{equation}
It follows immediately that $R$ can be expressed in terms of $Q$:
\begin{equation}
 \label{GammaXPTNumericalRelRQ}
   \begin{aligned}
   R=\sqrt{2E_3^2+2E_2^2-4m_\pi^2-Q^2}\,.
  \end{aligned}
\end{equation}
Furthermore, it is possible to express $V_\text{on}^2$ just in terms of the sum and the difference of $\vec{p}_2$ and $\vec{p}_3$, using $s=2m_\pi^2-2(p\cdot p_3)$, $t=2m_\pi^2-2(p\cdot p_2)$ and $u=4m_\pi^2-s-t$:
\begin{equation}
 \label{GammaXPTNumericalVonSumDiff}
   \begin{aligned}
   f_\pi^4\,V_\text{on}^2=7m_\pi^4 &+16m_\pi^2\left[E_p(E_3-E_2)-pQ\cos\theta_p\right]\\
    &+12\left[E_p(E_3-E_2)-pQ\cos\theta_p\right]^2\\
   &+4\left[E_p(E_3+E_2)-pR\cos\varphi\right]^2\,.
  \end{aligned}
\end{equation}
The second argument of $V_\text{on}^2(\cos\varphi_p,\cos\theta_p)$, is already fixed by \eqref{GammaXPTNumericalDeltaE}: $\cos\theta_p=x_0$. The first argument, $\cos\varphi_p$, can be related to $\cos\varphi$, by spherical trigonometry:
\begin{equation}
 \label{GammaXPTNumericalSpheric}
   \begin{aligned}
   \cos\varphi=\cos\theta_p\cos\Delta+\cos\varphi_p\sqrt{1-\cos^2\theta_p}\sqrt{1-\cos^2\Delta}\, .
  \end{aligned}
\end{equation}
Using its definition in \eqref{GammaXPTNumericalDefAnglesCo} and the relation \eqref{GammaXPTNumericalRelRQ}, $\cos\Delta$ is a function of the energies $E_2$, $E_3$ and $Q$ only:
\begin{equation}
 \label{GammaXPTNumericalDeltaQR}
   \begin{aligned}
   \cos\Delta=\frac{\vec{p}_3-\vec{p}_2}{Q}\cdot\frac{\vec{p}_3+\vec{p}_2}{R}=\frac{E_3^2-E_2^2}{QR(Q)}\,.
  \end{aligned}
\end{equation}
Inserting \eqref{GammaXPTNumericalDeltaQR} into \eqref{GammaXPTNumericalVonSumDiff}, $V_\text{on}^2$ becomes a quadratic polynomial in $\cos\varphi_p$:
\begin{equation}
\label{GammaXPTNumericalVonA}
  V_\text{on}^2(\cos\varphi_p)=A_0+A_1\cos\varphi_p+A_2\cos^2\varphi_p\,,
\end{equation}
with some complicated coefficient functions $A_i$ depending on $E_p$, $E_2,$, $E_3$, $m_\pi$, and $Q$. Now, we can easily carry out the integration over the azimuthal angle $\varphi_p$ from \eqref{EnergyEtaXPTformalIntegration}:
\begin{equation}
 \label{GammaXPTNumericalVonAIntegrated}
   \begin{aligned}
   \frac{1}{2\pi}\int_0^{2\pi}\dint\varphi_p\;\left(A_0+A_1\cos\varphi_p+A_2\cos^2\varphi_p\right)=A_0+\frac 12 A_2\,.
  \end{aligned}
\end{equation}
Consider next the two remaining momentum integrals $\int\dint^3p_2$ and $\int\dint^3p_3$ coming from Eq.~\eqref{SpectralWidthChPTFinal}. We have to ensure that the fixed value of $\cos\theta_p$ lies in the allowed range $|\cos\theta_p|=|x_0|\leq 1$. This constraint is implemented by the factor $\Theta(F(Q))$, with 
\begin{equation}
  \label{DefFunctionF}
  F(Q)=4p^2Q^2-\left[(E_3-E_2)(E_3-E_2+2E_p)-Q^2\right]^2\;,
\end{equation}
because $|x_0|\leq 1$ means
\begin{equation}
\label{GammaPhiFourNumericalCondCos}
  -1 \leq \frac{(E_3-E_2)(E_3-E_2+2E_p)-Q^2}{2p Q}\leq 1\;,
\end{equation}
which is equivalent to $F(Q)\geq 0$. As a function of $Q^2$, $F(Q)$ is just a concave-down parabola with $F(Q)\geq 0$ for $Q_1^2\leq Q^2\leq Q_2^2$ with roots $Q_1^2=(p-p_1)^2$ and $Q_2^2=(p+p_1)^2$. Since $Q\geq 0$ by its definition, we arrive at $Q_1=|p-p_1|$ and $Q_2=p+p_1$, the only two non-negative roots of $F(Q)$. At this stage, we arrive at the expression
\begin{equation}
\label{GammaXPTNumericalTwoMomenta}
\begin{aligned}
\int &\frac{\dint^3p_2}{E_2} \int\frac{\dint^3p_3}{E_3}\left(A_0+\frac 12 A_2\right)\frac{1}{4pQ}\;\Theta(F(Q))\\
   &=\frac{1}{4p}\int_{m_\pi}^\infty\dint E_3\int_{m_\pi}^{E_p+E_3-m_\pi}\dint E_2 \\
  &\hspace{0.3cm}\int\dint\Omega_2\int\dint\Omega_3\;\frac{p_2p_3}{Q}\left(A_0+\frac 12 A_2\right)\Theta(F(Q))\, .
  \end{aligned}
\end{equation}
The two remaining solid-angle integrals can be reduced to an integral over $Q$, using the relation
\begin{equation}
Q^2=p_2^2+p_3^2-2p_2p_3\cos\theta_{23}\;.
\end{equation}
Inspecting the coefficients $A_0$ and $A_2$, one observes that they consist of a few positive and negative powers of $Q^2$:
\begin{widetext}
\begin{equation}
 \label{GammaXPTNumericalTwoAnglesQ}
   \begin{aligned}
   &\frac{1}{4p} \int\dint\Omega_2\int\dint\Omega_3\;\frac{p_2p_3}{Q}\left(A_0+\frac 12 A_2\right)\Theta(F(Q))\\
   &=\frac{8\pi^2}{4p}\int_{|p_2-p_3|}^{p_2+p_3}\dint Q\;\left[5X_4Q^4+3X_2Q^2+X_0-X_{-2}Q^{-2}-3X_{-4}Q^{-4}\right]\Theta(F(Q))\\
   &=\frac{2\pi^2}{p}\left[X_4\left(Q_+^5-Q_-^5\right)+X_2\left(Q_+^3-Q_-^3\right)+X_0\left(Q_+-Q_-\right)+X_{-2}\left(\frac{1}{Q_+}-\frac{1}{Q_-}\right)+X_{-4}\left(\frac{1}{Q_+^3}-\frac{1}{Q_-^3}\right)\right].
  \end{aligned}
\end{equation}
\end{widetext}
Here the characteristic form of $X$ in Eq.~\eqref{DefFunctionXGeneral} with differences of powers of $Q_\pm$ has emerged. The step function $\Theta(F(Q))$ gives rise to a max/min pattern of the proper boundaries $Q_\pm$, which read
\begin{equation}
 \label{GammaXPTNumericalBoundariesQ}
   \begin{aligned}
    Q_-=\text{max}&\left\{|p_2-p_3|,|p-p_1|\right\},\\
    Q_+=\text{min}&\left\{p_2+p_3,p+p_1\right\}.
  \end{aligned}
\end{equation}
We combine the five coefficient functions $X_i(E_p,E_2,E_3;m_\pi)$ in the last line of Eq.~\eqref{GammaXPTNumericalTwoAnglesQ} to a new function $X(Q_+,Q_-)$:
\begin{equation}
  \label{DefFunctionXGeneral}
  X(Q_-,Q_+)=\sum_{i\in\{0,\pm 2,\pm 4\}} X_i\left(Q_+^{i+1}-Q_-^{i+1}\right).
\end{equation}
The explicit expressions for $X_i$ are the following:
\begin{widetext}
\begin{equation} 
 \label{GammaXPTNumericalXCoeffP4}
  X_4=\frac{7}{10}\,, \hspace{2cm} X_2=\frac 23\left[6m_\pi^2-E_p^2-4E_2^2-4E_3^2+E_pE_2-E_pE_3+7E_2E_3\right],
\end{equation}
\begin{equation}
\begin{aligned}
  X_0=\,& 15m_\pi^4+8m_\pi^2\left(3E_2E_3+E_pE_2-E_pE_3-E_p^2-2E_2^2-2E_3^2\right)\\
  &+2E_p\left(E_p-E_2+E_3\right)\left(5E_2^2+2E_2E_3+5E_3^2\right)+\left(E_2-E_3\right)^2\left(7E_2^2-4E_2E_3+7E_3^2\right),
\end{aligned}
\end{equation}

\begin{equation}
\begin{aligned}
  X_{-2} =\,& 4m_\pi^2\left(E_2-E_3\right)^2\left(2E_pE_2-2E_p^2-E_2^2-E_3^2-2E_pE_3\right) \\ 
  & +2E_p\left(E_2-E_3\right)^2\left(E_p-E_2+E_3\right) \left(7E_2^2+10E_2E_3+7E_3^2\right) +2\left(E_2-E_3\right)^4\left(2E_2^2+3E_2E_3+2E_3^2\right),
\end{aligned}
\end{equation}

\begin{equation}
\begin{aligned}
  \label{GammaXPTNumericalXCoeffM4}
  X_{-4}&=-\frac 12\left(E_2-E_3\right)^4\left(2E_p-E_2+E_3\right)^2\left(E_2+E_3\right)^2\,. \hspace{5cm}
\end{aligned}
\end{equation}
\end{widetext}
Putting all pieces together, we arrive at a well-conditioned double-integral representation for the spectral width:
\begin{equation}
 \label{GammaXPTNumericalBose}
   \begin{aligned}
    \Gamma(p) &=\frac{2n^{-1}(E_p)}{pE_pf_\pi^4\,(8\pi)^3}\int_{m_\pi}^\infty\dint E_3 \int_{m_\pi}^{E_p+E_3-m_\pi}\dint E_2 \\ 
     &\cdot X(Q_+,Q_-)n(E_1)n(E_2)[1+n(E_3)]\,\Theta(Q_+-Q_-)\,.
  \end{aligned}
\end{equation}
Alternatively, by using the identity
\begin{equation}
 \label{ChPTRelationBoseSinh}
  n(E_i)=\frac 12 \text{e}^{-\frac 12\beta E_i}\sinh^{-1}\left(\frac{\beta E_i}{2}\right),
\end{equation}
and energy conservation $E_1=E_p-E_2+E_3$, the spectral width can be expressed in the form given in Eq.~\eqref{GammaXPTNumericalSinh}. The additional factor $\Theta(Q_+-Q_-)$ in Eq.~\eqref{GammaXPTNumericalBose} must be introduced in order to restrict the double integral to the physical kinematical region in the $E_2E_3$-plane where the upper limit $Q_+$ is actually larger than the lower limit $Q_-$\,. At the same time this guarantees that the spectral width $\Gamma(p)$ is made up from strictly-positive contributions only.

\section{Discussion of ladder-diagram resummation for the skeleton expansion in \boldmath{$\chi$}PT}
\label{AppResum}
In order to approximate the four-point correlation function that enters $\Pi_\beta(\omega_n)$ in Eq. \eqref{spatIntThermGreen} we have performed a skeleton expansion \eqref{feySkeletonLONLO} from which we have taken only the one-loop term into account. A detailed analysis of this expansion for a real scalar theory with cubic and quartic self interactions has been given in \cite{JeonSkeleton1995, JeonYaffeSkeleton1996}. It is concluded there that for a consistent treatment of the shear viscosity $\eta$ at leading order in the small coupling constant $g\ll 1$, a resummation of all ladder diagrams needs to be performed:

\begin{quotation}
(\ldots) this means that higher loop diagrams can be just as important as the one-loop contribution if they are sufficiently infrared sensitive. \cite{JeonYaffeSkeleton1996}
\end{quotation}
This argument is based on the observation that in the limit of vanishing thermal width, $\Gamma\to 0$, the occurrence of pinched poles spoils the usual perturbative counting in powers of the small coupling $g$ (compare Fig. \ref{FigurePoles4x2}). The one-loop diagram in Eq. \eqref{feySkeletonLONLO} scales as $1/\Gamma$, whereas the two-loop diagram scales as $g^2/\Gamma^2$, and the $n$-loop ladder diagram scales as $g^{2n}/\Gamma^{n+1}$. Hence, there is a dimensionful scaling factor $g^2/\Gamma$ for every additional rung in the ladder-diagram expansion. Since the spectral width is $\Gamma\sim g^2$, all ladder diagrams are of the same order $\mathcal{O}(g^{-2})$, and therefore need to be resummed. According to Refs. \cite{Aarts03, Aarts04} this can be done in an efficient way by using a two-particle irreducible effective action.

Let us discuss to which extent $\chi$PT and scalar $g\phi^4$ theory differ in this respect. In the chiral limit, $m_\pi=0$, the pion-pion interaction is purely of derivative type, i.e. proportional to $p^2$. In such a situation, the infrared singular $1/\Gamma(p)$ terms resulting from the nearly pinching poles are compensated by momentum-dependent factors in the numerator. Inspecting the chiral Lagrangian \eqref{ChPTLagrangian2pi4} one can identify the dimensionless coupling $g\,\hat{=}\,p^2/2f_\pi^2$. The additional factor appearing at each higher order in the ladder-diagram expansion is $g^2/\Gamma(p)=p^4/(4f_\pi^4\Gamma(p))$, which vanishes in the infrared limit, $p\to 0$. However, the additional chiral-symmetry breaking mass term in the chiral Lagrangian gives rise to a pion-pion interaction analogous to the vertex in $g\phi^4$ theory, with $g\,\hat{=}\,m_\pi^2/8f_\pi^2$. Following the scaling arguments of \cite{JeonSkeleton1995} this feature may require the resummation of ladder diagrams.

In general, $\chi$PT is far less infrared sensitive than other bosonic field theories. While a detailed analysis needs yet to be performed, one expects that the numerical consequences of such a resummation may be less important than in $g\phi^4$ theory. Let us finally note that $\chi$PT is only applicable for low temperatures $T\lesssim m_\pi=140\;\text{MeV}$. In that temperature range, thermal corrections to the pion mass are smaller than $5\%$ and therefore negligible \cite{Toublan1997,Kaiser1999}. For instance, this is manifest in the absence of a Linde problem \cite{Kapusta} in $\chi$PT.

\section*{Acknowledgments}
We thank G. Aarts for useful remarks and references concerning the ladder resummation in scalar field theory. This work is partially supported by the German Bundesministerium f\"{u}r Bildung und Forschung (BMBF), the TUM Graduate School (TUM-GS), and the DFG Cluster of Excellence ``Origin and Structure of the Universe''.

\end{document}